\documentclass[aps,prd,showpacs,floatfix,preprintnumbers,groupedaddress]{revtex4}
\usepackage{graphicx}
\usepackage{psfrag}
\def\lsim{\raise0.3ex\hbox{$\;<$\kern-0.75em\raise-1.1ex\hbox{$\sim\;$}}}
\def\gsim{\raise0.3ex\hbox{$\;>$\kern-0.75em\raise-1.1ex\hbox{$\sim\;$}}}

\def\vev#1{\left\langle #1\right\rangle}

\def\hbar{\hspace{0pt}\raisebox{1pt}{$-$} \hspace{-7pt} h}

\newcommand{\be}{\begin{equation}}
\newcommand{\ee}{\end{equation}}
\newcommand{\bd}{\begin{displaymath}}
\newcommand{\ed}{\end{displaymath}}
\newcommand{\bea}{\begin{eqnarray}}
\newcommand{\eea}{\end{eqnarray}}

\newcommand {\ignore}[1]{}

\def\10{SO(10)}
\def\321{SU(3) $\otimes$ SU(2) $\otimes$ U(1) }

\newcommand{\AddrAHEP}{%
  AHEP Group, Institut de F\'{\i}sica Corpuscular --
  C.S.I.C./Universitat de Val{\`e}ncia \\
  Edificio Institutos de Paterna, Apartado 22085, E--46071 Valencia, Spain}
\newcommand{\AddrLisb}{%
 Departamento de F\'\i sica and CFTP, Instituto Superior T\'ecnico\\
          Avenida Rovisco Pais 1, 1049-001 Lisboa, Portugal }

\begin{document}

\preprint{IFIC/07-37}

\title{Fermion masses, Leptogenesis and Supersymmetric \10 Unification}
\date{\today}

\author{J.~C.~Rom\~ao} \email{jorge.romao@ist.utl.pt}\affiliation{\AddrLisb}
\author{M.~A.~T\'ortola}\email{mariam@cftp.ist.utl.pt}\affiliation{\AddrLisb}
\author{M.~Hirsch} \email{mahirsch@ific.uv.es} \affiliation{\AddrAHEP}
\author{J.~W.~F.~Valle} \email{valle@ific.uv.es} \affiliation{\AddrAHEP}

\begin{abstract}

 Current neutrino oscillation data indicate the existence of two
 large lepton mixing angles, while Kobayashi-Maskawa matrix elements
 are all small.  Here we show how supersymmetric \10 with extra
 chiral singlets can easily reconcile large lepton mixing angles with
 small quark mixing angles within the framework of the successful
 Fritzsch ansatz. Moreover we show how this is fully consistent with
 the thermal leptogenesis scenario, avoiding the so-called gravitino
 problem. A sizeable asymmetry can be generated at scales as low
 as possible within the leptogenesis mechanism. We present our
 results in terms of the leptonic CP violation parameter that
 characterizes neutrino oscillations. 

\end{abstract}

\maketitle

\section{Introduction}

An endemic difficulty in unified gauge models is the understanding of
fermion masses. Thanks to the brilliant series of experiments which
led to the discovery of neutrino oscillations over the past ten years
or so, we now have an improved knowledge of neutrino masses and
mixings~\cite{Maltoni:2004ei}.
The maximal atmospheric mixing angle and the large solar mixing angle
both came as a surprise, at odds with naive unification--based
expectations that lepton and quark mixings are similar.
Indeed, although not mandatory, within unification models a similar
structure for quark and lepton mixing is far easier to account for
than what is currently established by observation.
Several approaches have been considered in the literature to
circumvent this problem, involving various types of extensions of the
multiplet
content~\cite{Babu:1998wi,bertolini2004fms,Dermisek:2005ij,Altarelli:2004za,Nasri:2004rm}.

A virtue of seesaw models is that they bring the possibility of
decoupling the angles in the quark and lepton sectors by having
additional Majorana-type terms for the latter~\cite{schechter:1980gr}.
Moreover, seesaw models may account for the observed cosmological
baryon excess through the elegant mechanism called
leptogenesis~\cite{Fukugita:1986hr}.
However in minimal supergravity models, with $m_{3/2} \sim$ 100 GeV to
10 TeV the supersymmetric type-I seesaw scenario leads to an
overproduction of gravitinos after
inflation~\cite{Khlopov:1984pf,Ellis:1984eq} unless the reheat
temperature is restricted to be much lower than that required for
successful leptogenesis~\cite{Kawasaki:2004qu}.
One possible way out is resonant leptogenesis~\cite{Pilaftsis:2005rv}
as indeed suggested in Ref.~\cite{Akhmedov:2003dg}. Another
alternative has been considered in Ref.~\cite{Farzan:2005ez} requires
going beyond the minimal seesaw by adding a small R-parity violating
term in the superpotential.

Here we consider an alternative approach to supersymmetric \10
unification that involves mainly an extension of the lepton sector and
a novel realization of the seesaw mechanism already discussed
previously~\cite{Hirsch:2006ft,Malinsky:2005bi}.
We show how the observed structure of lepton masses and mixings fits
together, thanks to the presence of the extra states, with the
requirements of a predictive pattern of quark mixing, encoded within
the successful Fritzsch ansatz~\cite{Fritzsch:1977vd}.
Moreover, we show explicitly how such extended supersymmetric \10
seesaw scheme provides a realistic scenario for thermal leptogenesis,
avoiding the so-called gravitino problem.
Our approach is phenomenological in sipirit. Hence we do not seek to
derive a full-fledged unified theory of flavour incorporating a
specific symmetry, such as
$A_4$~\cite{babu:2002dz,Altarelli:2005yx,Hirsch:2006je,Hirsch:2005mc,Hirsch:2007kh}
to the unified gauge group level. 

The paper is organized as follows. In Sec.~\ref{sec:minimal-type-i} we
discuss the limitations of minimal supersymmetric type-I seesaw in
naturally providing a framework for thermal leptogenesis in agreement
with bounds on the reheat temperature after inflation, in
Sec.~\ref{sec:model} we recall the basic features of the considered
model, while ansatzes for the coupling matrices are discussed in
Sec.~\ref{sec:ansatz-coupl-matr}. In Sec. \ref{sec:washout} 
we discuss the calculation of the final baryon number asymmetry 
from the decay asymmetry of the lightest singlet. There we discuss 
how we can use results obtained previously on production and washout 
factors in the seesaw type-I case also in our extended seesaw model. 
We show that leptogenesis scenario
is consistent both with the reheat temperature constraint as well as
with fitting the lepton and quark mixing angles within the successful
Fritzsch ansatz.  

\section{Minimal supersymmetric seesaw leptogenesis}
\label{sec:minimal-type-i}

In the most general seesaw model the Dirac fermion mass terms arise
from the couplings of the {\bf 16} with the {\bf 10} and {\bf 126},
while the diagonal entries in the seesaw neutrino mass matrix arise
only from the couplings of the {\bf 16} with the {\bf 126}.  The mass
matrix expressed in the basis is $\nu_{L}$, $\nu^{c}_{L}$ is given
as~\cite{Valle:2006vb},
\begin{equation}
\label{ss-matrix} {\mathcal M_\nu} = \left(\begin{array}{cc}
    Y_L \vev{\Delta_L} & Y \vev{\Phi} \\
    {Y}^{T} \vev{\Phi}  & Y_{R} \vev{\Delta_R} \\
\end{array}\right)\,,
\end{equation}
where the {\bf 16} denotes each chiral matter generation while the
{\bf 10} and {\bf 126} are Higgs-type chiral multiplets.
Here \(Y\), \(Y_L\) and \(Y_R\) denote the Yukawa couplings of the
{\bf 10} and {\bf 126}, respectively. These are all symmetric in
flavour, the symmetry of \(Y_L,~Y_R\) results from the Pauli
principle, while that of \(Y\) follows from \10.

Minimization of the Higgs scalar potential leads to a vev seesaw
relation
 \begin{equation}
  \vev{\Delta_{L}} \vev{\Delta_{R}} \sim \vev{\Phi}^2~,
 \label{eq:vev-seesaw}
 \end{equation}
 consistent with the desired vev hierarchy $ \vev{\Delta_{R}} \gg
 \vev{\Phi} \gg \vev{\Delta_{L}}$,  where $\vev{\Phi}$ is the standard
 vev and $\vev{\Delta_{L,R}}$ are the vacuum expectation values (vevs)
 giving rise to the Majorana terms.

 Before we display our proposed model it is useful to consider the
 case of the Minimal type-I seesaw~\cite{Minkowski:1977sc}. This
 consists in taking $\vev{\Delta_{L}} \to 0$ in the above equations.
 Although not generally consistent with \10 in the presence of the
 {\bf 126}, this approximation is suitable for our purposes as the
 extension we will be discussing is tripletless, the {\bf 126} being
 replaced by {\bf 16}'s.

 The matrix \(\mathcal{M_\nu}\) is diagonalized by performing a
 a \(6\times6\) unitary transformation \(U_\nu\),
\begin{equation}
\label{eq:light-nu}
   \nu_i = \sum_{a=1}^{6}(U_\nu)_{ia} n_a ~~~~~~~~
U_\nu^T {\mathcal M_\nu} U_\nu = \mathrm{diag}(m_i,M_i) ,
\end{equation}
whose form is given explicitly as a perturbation series in
Ref.~\cite{schechter:1982cv}. 
The effective mass matrix of the three light neutrinos is given as
\begin{equation}
  \label{eq:ss-formula}
  m_{\nu} \approx  - Y {Y_R}^{-1} {Y}^T \frac{{\vev{\Phi}}^2}{\vev{\Delta_R}}
\end{equation}
The smallness of light neutrino masses follows dynamically from
Eq.~(\ref{eq:ss-formula}), since $\vev{\Delta_R}$ is large.

In order to reproduce the observed pattern of quark masses and CP
violation we assume the complex matrix $Y$ to have the form
\begin{equation}
  \label{eq:fritsch-1}
  Y=\left(
    \begin{array}{ccc}
      0 & a & 0 \\[+1mm]
      a^* & b & c \\[+1mm]
      0 & c^* & d \\[+1mm]
    \end{array}
\right)
\end{equation}
with $a,c$ complex and $b,d$ real. Notice that the large observed
lepton mixing angles can always be reconciled with the small quark
mixing angles thanks to the presence of the coupling matrices
$Y_{L,R}$ which exist only for neutrinos~(in the limit of unbroken
D-parity one has \(Y_L=Y_R\)).
However, in a unified theory where a flavour symmetry is assumed to
predict the lepton mixing angles, it becomes a real challenge to
account naturally also for the quark mixing angles.

\begin{figure}[t]
\centering
\includegraphics[height=15cm, angle=-90]{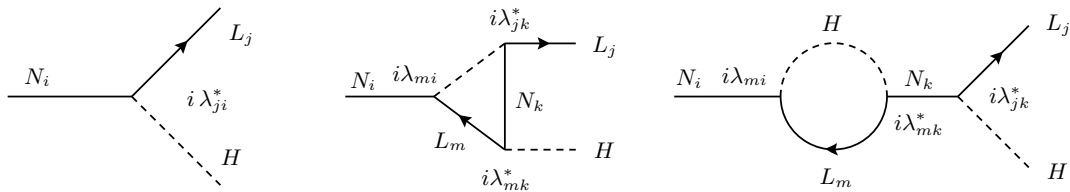}
\caption{\label{fig:tl+oneloop} Diagrams contributing to generate a
  lepton asymmetry of the universe.}
\end{figure}

Now we turn to cosmology.  One of the attractive features of seesaw
models is that they open the possibility of accounting for the
observed cosmological matter-antimatter asymmetry in the Universe
through the leptogenesis mechanism~\cite{Fukugita:1986hr}.
This requires the out-of-equilibrium decays of the heavy
``right-handed'' neutrinos $N_i$ (see Fig.~\ref{fig:tl+oneloop}) to
take place before the electroweak phase transition, and the presence
of CP violation in the lepton sector. The tree level and one loop
diagrams for $N_i$ decay that interfere in order to generate a lepton
asymmetry of the universe are shown.

These decays generate the asymmetry,
\begin{equation}
  \label{eq:37}
  \epsilon_i = \frac{\Gamma_{N_i L} -\Gamma_{N_i\overline{L}}}
{\Gamma_{N_i L} + \Gamma_{N_i\overline{L}}}
\end{equation}
Here we recall the estimate of such asymmetry, as given in
Ref.\cite{covi:1996wh}. The relevant amplitudes follow from the
initial Lagrangian
\begin{equation}
  \label{eq:24}
  \mathcal{L}= \lambda_{ij} \overline{N}_i P_L L_j H + {\rm h.c.}.
\end{equation}
Here it is sufficient just to sketch their structure. At the tree
level we have
\begin{equation}
  \label{eq:38}
  \mathcal{M}^0_{N L} = i \lambda_{ji}^*, \qquad
  \mathcal{M}^0_{N \overline{L}} = i \lambda_{ji},
\end{equation}
while for the one-loop it is only necessary to evaluate the vertex
contribution, 
\begin{equation}
  \label{eq:39}
  \mathcal{M}^1_{N L} = i^7\, f\, M_k  \lambda_{jk}^*  \lambda_{mk}^* 
  \lambda_{mi}, \qquad
  \mathcal{M}^1_{N \overline{L}} = i^7  f\, M_k  \lambda_{jk}  \lambda_{mk} 
  \lambda_{mi}^*
\end{equation}
where $f$ is some complex loop function. Now tree level plus one loop
will be proportional to
\begin{eqnarray}
  \label{eq:40}
|\mathcal{M}^0_{N L} +  \mathcal{M}^1_{N L}|^2 &\simeq&
 |\mathcal{M}^0_{N L}|^2 + \mathcal{M}^0_{N L} \mathcal{M}^{1 *}_{N L} +
\mathcal{M}^{0 *}_{N L} \mathcal{M}^1_{N L} 
\end{eqnarray}
so that the asymmetry is
\begin{equation}
  \label{eq:41}
  \epsilon_i \propto \frac{\mathcal{M}^0_{N L} \mathcal{M}^{1 *}_{N L} +
\mathcal{M}^{0 *}_{N L} \mathcal{M}^1_{N L} -\mathcal{M}^0_{N \overline{L}}
\mathcal{M}^{1 *}_{N \overline{L}} -
\mathcal{M}^{0 *}_{N \overline{L}} \mathcal{M}^1_{N \overline{L}} }
{|\mathcal{M}^0_{N L}|^2}
\end{equation}
Now using Eq.~(\ref{eq:38}) and Eq.~(\ref{eq:39}) we can write
\begin{equation}
  \label{eq:42}
  \epsilon_i \propto {\rm Im}(f) \frac{\sum_{k,j,m} {\rm Im} \left(\lambda_{ji}^*
  \lambda_{mi}^*  \lambda_{mk}  \lambda_{jk} \right) M_k}{\sum_j \lambda_{k i}^*
\lambda_{k i}}
\end{equation}
Finally we get, in the case of hierarchical right-handed neutrinos,
\begin{equation}
  \label{eq:43}
  \epsilon_1 = -\frac{3 }{16 \pi} M_1 \displaystyle
 \frac{\displaystyle \sum_{k\not=1,j,m} {\rm Im} \left(\lambda_{j 1}^*
  \lambda_{m 1}^*  \lambda_{mk}  \lambda_{jk} \right)
\frac{1}{M_k}}{\displaystyle  \sum_k \lambda_{j 1}^* \lambda_{j 1}}
\end{equation}
Note that in this limit the contribution of the self energy has the
same dependence on the couplings and is already included.  The lepton
(or B-L) asymmetry thus produced then gets converted, through
sphaleron processes, into the baryon asymmetry~\cite{Fukugita:1986hr},
observed to be $\mathcal O(10^{-10})$.
In order to provide an acceptable framework for leptogenesis, and
taking into account the presence of washout effects, the asymmetry we
need is $\epsilon_1 \sim \mathcal O(10^{-7})$ or larger, for the same
values of parameters that reproduce the observed small neutrino
masses, Eq.~(\ref{eq:ss-formula}).
In order to generate this asymmetry thermally one requires the
Universe to reheat after inflation to a very high
temperature~\cite{Buchmuller:2004nz,Davidson:2002qv,Hamaguchi:2001gw,Barbieri:1999ma}
\begin{equation}
  \label{eq:RHreq}
T_R > 2 \times 10^9~\mathrm{GeV}
\end{equation}
Such large scale leads to an overproduction of cosmological
gravitinos.
In minimal supergravity models, with $m_{3/2} \sim$ 100 GeV to 10 TeV
gravitinos are not stable, decaying during or after Big Bang
Nucleosynthesis (BBN). Their rate of production can be so large that
subsequent gravitino decays completely change the standard BBN
scenario.
Since the abundance of gravitinos is proportional to $T_R$ such
``gravitino crisis'' can only be prevented by requiring a low enough
reheat temperature $T_R$ after
inflation~\cite{Khlopov:1984pf,Ellis:1984eq}.
A recent detailed analysis gives a stringent upper bound 
\begin{equation}
  \label{eq:RH}
  T_R \lsim  10^6~\mathrm{GeV}
\end{equation}
when the gravitino has hadronic decay modes~\cite{Kawasaki:2004qu}.
Therefore, thermal leptogenesis seems difficult to reconcile with low
energy supersymmetry if gravitino masses lie in the range suggested by
the simplest minimal supergravity models. One possible way out is to
have resonant leptogenesis~\cite{Pilaftsis:2005rv} as suggested in
Ref.~\cite{Akhmedov:2003dg}. Another alternative considered in
Ref.~\cite{Farzan:2005ez} requires going beyond the minimal seesaw by
adding a small R-parity violating term in the superpotential.
In the following we discuss quantitatively the alternative suggestion
made in~\cite{Hirsch:2006ft} in the context of the extended
supersymmetric seesaw scheme.


\section{The new extended seesaw model}
\label{sec:model}


For definiteness we work in the context of the supersymmetric \10
unified model considered in Ref.\cite{Malinsky:2005bi,Hirsch:2006ft}.
The gauge symmetry and D-parity are broken by {\bf 45} and {\bf 210}
multiplets at the unification scale.
The B-L symmetry breaks at lower scale thanks to expectation values of
the $\chi$ fields in the {\bf 16}, instead of the more familiar left
and right triplets present in {\bf 126} which lead to the standard
seesaw mechanism.  Gauge couplings unification can not fix the B-L
breaking scale, which can be relatively low, as shown in
Ref.~\cite{Malinsky:2005bi}. The possibility of a low B-L breaking
scale also fits with the observed neutrino masses.  The relevant
Yukawa couplings leading to neutrino masses are
\begin{equation}
\label{eq:lag}
-{\cal L}_Y = Y_{ij} {\nu^c}_{iL} \nu_{jL} \phi + F_{ij} \nu_{iL}
S_j \chi_L + \tilde{F}_{ij}  {\nu^c}_{iL} S_j \chi_R + \frac{1}{2}
M_\Sigma \Sigma \Sigma.
\end{equation}
Note that a direct Majorana mass term for the singlet fields $S_i$ is
forbidden by an additional imposed $U(1)_G$ symmetry and the fact that
the only singlet scalar present ($\sigma$) is odd under D-parity,
while $S_i S_j$ are even under D-parity. For the same reason, $\sigma$
cannot couple to $\Sigma$, but a bare mass is
allowed~\cite{Hirsch:2006ft}. As we will show leptogenesis allows this
mass to be of the order of TeV.
We also introduce a soft term breaking $U(1)_G$, which
allows mixing between the scalar components of these fields $\Sigma
S_i$,
\begin{equation}
-{\cal L}_{\not G} = \Delta_i \Sigma S_i,
\end{equation}
This will then give  a $10 \times 10$ neutrino mass matrix, in the
basis ($\nu_i$, $\Sigma$, $\nu^c_{iL}$, $S_i$):
\begin{equation}
\label{eq:nu-mass-mat}
\mathcal{M_\nu} =\left(
\begin{array}{cccc}
0 & 0 &Y v & F v_{L}  \\
0 & M_{\Sigma} & 0 & \Delta_{}^{T} \\
Y^{T} v & 0 & 0 & \tilde{F} v_{R} \\
F^{T} v_{L} & \Delta_{} & \tilde{F}^{T}v_{R} & 0 \\
\end{array}
\right)
\end{equation}
where, $v = \vev{\phi}$, $v_L = \vev{\chi_L}$ and $v_R = \vev{\chi_R}$
are the vevs for the fields $\phi$, $\chi_L$ and $\chi_R$ respectively
and $ \Delta_{}$ is the $U(1)_G$ breaking entry.  Besides the three
light neutrinos, this mass matrix will give two heavy states which are
dominantly the right-handed neutrino ${\nu^c}_{iL}$ and the singlets
$S_i$, and a lighter state $\Sigma$.
Notice that we will not impose any specific family symmetry.
In Sec.~\ref{sec:hirerarchicalY} we will, however, consider the case
where the $Y_{ij}$ are restricted to follow the Fritzsch texture for
the quark sector.

\subsection{Light  neutrinos}

Using the seesaw diagonalization prescription given in
Ref.~\cite{schechter:1982cv} we obtain the effective left-handed light
neutrino mass matrix as~\cite{Hirsch:2006ft},
\begin{equation}
\label{eq:mnua}
m_{\nu} = - \frac{1}{M_{\Sigma}}G G^{T}-\left[Y (F
\tilde{F}^{-1})^{T}+(F \tilde{F}^{-1})Y^{T}\right]\frac{v v_{L}}{v_{R}}
\end{equation}
where $G \equiv Y (\tilde{F}^{-1})^{T} \Delta \frac{v}{v_{R}}$.
Since the $D-$parity breaking scale is much higher than the scale at
which the left-right symmetry breaks, and this in turn is higher than
the electroweak symmetry breaking scale, one has the ``vev-seesaw''
relation
\begin{equation}
\label{eq:vev-seesaw-2}
v_L \propto {v_R v \over M_X },
\end{equation}
where $ M_X$ is determined by the \10 breaking vevs. In the following
we will take this proportionality constant equal to one and therefore
write Eq.~(\ref{eq:mnua}) as
\begin{equation}
\label{eq:mnu}
m_{\nu} = - \frac{1}{M_{\Sigma}}G G^{T}-\left[Y (F
\tilde{F}^{-1})^{T}+(F \tilde{F}^{-1})Y^{T}\right]\frac{v^2}{M_X}
\end{equation}
Note that the neutrino masses are suppressed by the unification scale
$M_X$ instead of the B-L scale. As noted in \cite{Malinsky:2005bi} we
need two pairs of {\bf 16 $\overline{\bf 16}$} in order to boost up
the neutrino mass scale and bring in an independent flavour structure
beyond that of the charged Dirac couplings, constrained by the quark
sector phenomenology.

Now we want to extract as much information from the low energy
neutrino data as possible. The matrix $m_{\nu}$ can be diagonalized
as,
\begin{equation}
  \label{eq:1}
  U^T m_{\nu} U = {\rm diag}(m_{\nu_1},m_{\nu_2},m_{\nu_3})
\end{equation}
where we use the standard parameterization from
Ref.\cite{schechter:1980gr} and now adopted by the PDG
\begin{equation}
  \label{eq:2}
  U=\left(
    \begin{array}{ccc}
      c_{12} c_{13}& s_{12} c_{13} & s_{13} e^{-i\delta} \\
      -s_{12} c_{23} -c_{12} s_{23} s_{13} e^{i\delta}
      & c_{12} c_{23} -s_{12} s_{23} s_{13} e^{i\delta}  & s_{23} c_{13} \\
      s_{12} s_{23} -c_{12} c_{23} s_{13} e^{i\delta}
      & -c_{12} s_{23} -s_{12} c_{23} s_{13} e^{i\delta}  & c_{23} c_{13} \\
    \end{array}
\right)
\end{equation}
The values of $U$ and ${\rm diag}(m_{\nu_1},m_{\nu_2},m_{\nu_3})$ are
known to some degree from experiment and we want to turn this
information into the parameters of the theory, defined by
Eq.~(\ref{eq:lag}). Of course we have too little experimental
information, three angles and two mass differences, to be able to
reconstruct in full the three matrices $Y, F, \tilde{F}$ and the
vector $\Delta$.

\subsection{Heavy neutrinos}

Let us now turn to the discussion of the heavy neutrinos. The
following terms in the Lagrangian,
\begin{eqnarray}
  \label{eq:17}
  \mathcal{-L} &=& \frac{1}{2}\left( 
  2 \tilde{F}_{ij}  {\nu^c}_{iL} S_j v_R +
 M_\Sigma  \Sigma \Sigma + 2 \Delta_i \Sigma S_i \right)\nonumber \\
&=&\frac{1}{2}\left(
v_R\, {\nu^c}_{L}^T \tilde{F} S +
v_R\, S^T \tilde{F}^T {\nu^c}_{L} 
+\Delta^T \Sigma S + \Delta S^T \Sigma  
+M_\Sigma  \Sigma \Sigma
\right)\nonumber \\
&=& \frac{1}{2} N_i M_{N ij} N_j
\end{eqnarray}
lead to a $7\times7$ mass matrix in the basis $N=$($\Sigma$,
$\nu^c_{iL}$, $S_i$):
\begin{equation}
  \label{eq:16}
  M_{N}=\left(
    \begin{array}{ccc}
      M_{\Sigma} & 0 & \Delta^T\\[+2mm]
      0 & 0 & \tilde{F} v_R\\[+2mm]
      \Delta & \tilde{F}^T v_R & 0\\[+2mm]
    \end{array}
\right)
\end{equation}
which determines the masses of the heavy neutrinos.
Now in order to get the states with physical masses we diagonalize
this mass matrix in three steps. We start by diagonalizing the matrix
$\tilde{F}$. We define
\begin{equation}
  \label{eq:18}
  U_N^T \tilde{F} U_S \ v_R= {\rm diag}(M_1,M_2,M_3) \equiv M_D
\end{equation}
where $U_N$ and $U_S$ are unitary matrices. Then we define new fields
\begin{equation}
  \label{eq:19}
  \nu'^c = U_N^{-1} \nu^c, \quad S'=U_S^{-1} S
\end{equation}
so that in the basis $N'=(\Sigma, \nu'^c, S')$ we have
\begin{equation}
  \label{eq:20}
  \mathcal{-L} =\frac{1}{2} N'^T M'_N N'
\end{equation}
with
\begin{equation}
  \label{eq:21}
  M'_N= R_{NS}^T M_N R_{NS} =\left(
    \begin{array}{ccc}
      M_{\Sigma} & 0 & \Delta^T U_S \\[+1mm]
      0 & 0 & M_D \\[+1mm]
      U_S^T \Delta  & M_D & 0 \\[+1mm]
    \end{array}
\right)
\end{equation}
and
\begin{equation}
  \label{eq:22}
  R_{NS}=\left(
    \begin{array}{ccc}
      1 & 0 & 0\\[+1mm]
      0 & U_N & 0\\[+1mm]
      0 & 0 & U_S\\[+1mm]
    \end{array}
\right)
\end{equation}

Now we rotate the fields $\nu'^c$ and $ S'$ to obtain the almost
degenerate states $N_{\pm}=\frac{1}{\sqrt{2}} (\nu'^c \pm S')$. This is
done by defining a new matrix 
\begin{equation}
  \label{eq:23}
  R=\left(
    \begin{array}{ccc}
     1 & 0 & 0 \\ [+1mm]
     0 & \displaystyle \frac{1}{\sqrt{2}} &  
     \displaystyle \frac{1}{\sqrt{2}}\\ [+1mm]
     0 & -\displaystyle \frac{1}{\sqrt{2}} & 
     \displaystyle \frac{1}{\sqrt{2}} \\ [+1mm]
    \end{array}
\right)
\end{equation}
that obeys
\begin{equation}
  \label{eq:25}
   \mathcal{-L} =\frac{1}{2} N''^T M''_N N''
\end{equation}
with $N' = R N''$, $N''=(\Sigma, N_-, N_+)$  and 
\begin{equation}
  \label{eq:26}
  M''_N = R^T M'_N R = \left(
    \begin{array}{ccc}
            M_{\Sigma} &\displaystyle -\frac{1}{\sqrt{2}} \Delta^T U_S & 
            \displaystyle \frac{1}{\sqrt{2}}  \Delta^T U_S \\[+2mm]
           \displaystyle -\frac{1}{\sqrt{2}} U_S^T \Delta & -M_D & 0 \\[+2mm]
           \displaystyle \frac{1}{\sqrt{2}}U_S^T \Delta  & 0 & M_D \\[+2mm]
    \end{array}
\right)
\end{equation}

We now have to perform the final rotation in order to get the physical
states for these heavy neutrinos. As we have seen in the discussion of
the light neutrinos, the amount of $G$-violation must be small,
therefore Eq. (\ref{eq:26}) can be approximately diagonalized using
the techniques of Ref.\cite{schechter:1982cv}. We obtain
\begin{equation}
  \label{eq:27}
  \mathcal{-L} =\frac{1}{2} N'''^T M'''_N N'''
\end{equation}
with $N'''=(\Sigma', N'_{-},N'_{+})$, $N''= U_{RH} N'''$ where
\begin{equation}
  \label{eq:28}
  M'''_N=U_{RH}^T M''_N U_{RH} = \left(
    \begin{array}{ccc}
      M_{\Sigma} & 0 & 0 \\[+1mm]
      0 & -M_D & 0 \\[+1mm]
      0 & 0 & M_D \\[+1mm]
    \end{array}
\right)
\end{equation}
and
\begin{equation}
  \label{eq:29}
  U_{RH}=\left(
    \begin{array}{ccc}
      1 &\displaystyle \frac{1}{\sqrt{2}} \Delta^T
      \left(U_S^{-1}\right)^T M_D^{-1} &
      \displaystyle \frac{1}{\sqrt{2}} \Delta^T
      \left(U_S^{-1}\right)^T M_D^{-1} \\[+2mm]
      \displaystyle -\frac{1}{\sqrt{2}} M_D^{-1} U_S^T \Delta
      & 1 & 0\\[+2mm]
      \displaystyle -\frac{1}{\sqrt{2}} M_D^{-1} U_S^T \Delta
      & 0 & 1\\[+2mm]
    \end{array}
\right)
\end{equation}
If we assume that the eigenvalues of $\tilde{F}$ are hierarchical we
get also a hierarchical spectrum for the heavy states, $\Sigma',
N'_{-}, N'_{+}$, namely, $ |M_{\Sigma}| \ll |M_1| \ll |M_2| \ll
|M_3|$.
In order to evaluate the asymmetry generated we must evaluate the
couplings of the mass eigenstates. For this we notice that
\begin{equation}
  \label{eq:31}
  N= R_{NS} R  U_{RH} N'''
\end{equation}
and a simple calculation gives
\begin{equation}
  \label{eq:32}
  R_{NS} R  U_{RH}=\left(
    \begin{array}{ccc}
      1 & \displaystyle \frac{1}{\sqrt{2}} \Delta^T
      \left(U_S^{-1}\right)^T M_D^{-1} &
      \displaystyle \frac{1}{\sqrt{2}} \Delta^T
      \left(U_S^{-1}\right)^T M_D^{-1}\\[+2mm]
      -  U_N  M_D^{-1} U_S^T \Delta &
      \displaystyle \frac{1}{\sqrt{2}} U_N &
      \displaystyle \frac{1}{\sqrt{2}} U_N \\[+2mm]
      0  &  \displaystyle - \frac{1}{\sqrt{2}} U_S &
      \displaystyle \frac{1}{\sqrt{2}} U_S \\[+2mm]
    \end{array}
\right)
\end{equation}
which allows us to write
\begin{eqnarray}
  \label{eq:33}
  \Sigma &=& \Sigma' +\frac{1}{\sqrt{2}} \Delta^T
      \left(U_S^{-1}\right)^T M_D^{-1} \ N'_{-} +
      \frac{1}{\sqrt{2}} \Delta^T
      \left(U_S^{-1}\right)^T M_D^{-1} \ N'_{+} \nonumber \\[+2mm]
\nu^c &=& -\left(\tilde{F}^{-1}\right)^T \frac{\Delta}{v_R}\Sigma' 
+ \frac{1}{\sqrt{2}} U_N N'_{-} + 
 \frac{1}{\sqrt{2}} U_N N'_{+} \\[+2mm]
S&=&- \frac{1}{\sqrt{2}} U_S\ N'_{-} +\frac{1}{\sqrt{2}} U_S\ N'_{-}
\nonumber 
\end{eqnarray}
where we have used
\begin{equation}
  \label{eq:34}
  \left(\tilde{F}^{-1}\right)^T =  U_N  M_D^{-1} U_S^T\, v_R
\end{equation}

With this we can rewrite the relevant part of the Lagrangian of
Eq.~(\ref{eq:lag}) in terms of the eigenstates (we drop the primes from
now on),
\begin{equation}
  \label{eq:35}
  {\cal L}_Y = Y_{\Sigma i}\, L_i\, H\, \Sigma + Y_{\pm ij}\, N_{\pm i}\, L_j\, H
  + {\rm h.c.}   
\end{equation}
where
\begin{equation}
  \label{eq:36}
  Y_{\Sigma } = \alpha_H Y \left(\tilde{F}^{-1}\right)^T
  \frac{\Delta}{v_R}, \quad
  Y_{\pm ij} = - \left(\frac{1}{\sqrt{2}} 
\alpha_H \left(U_N^T\, Y\right)_{ij} \pm \alpha_\chi^H 
\left( F\, U_S\right)_{ji} \right)
\end{equation}
where $\alpha_H$ denotes the projection of the relevant light MSSM
Higgs doublet $h$ into the directions of the defining Higgs doublets
living in $H \in 10_H$, and $\alpha_{\chi^{k}}^H$ are the projections
of the light MSSM-like Higgs doublet onto the defining Higgs doublets
in the $\overline{16}_{H}^{k}$ and $M_{1}$ is the mass of the $N_{\pm
1}$.

\subsection{Calculation of the Asymmetry}

\begin{figure}[b!] 
\centering
\includegraphics[height=3cm]{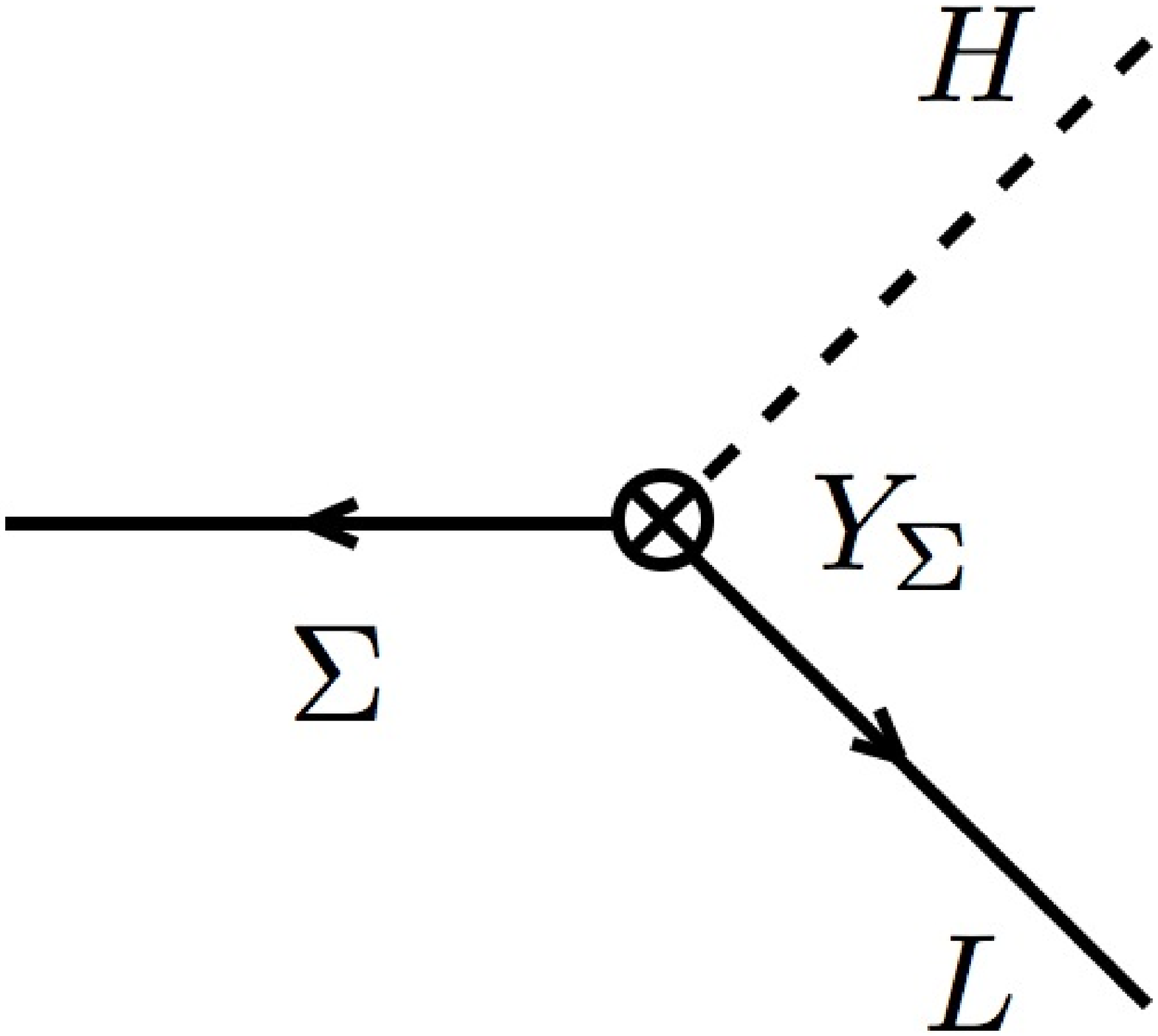}~~~~~~~
\includegraphics[height=3cm]{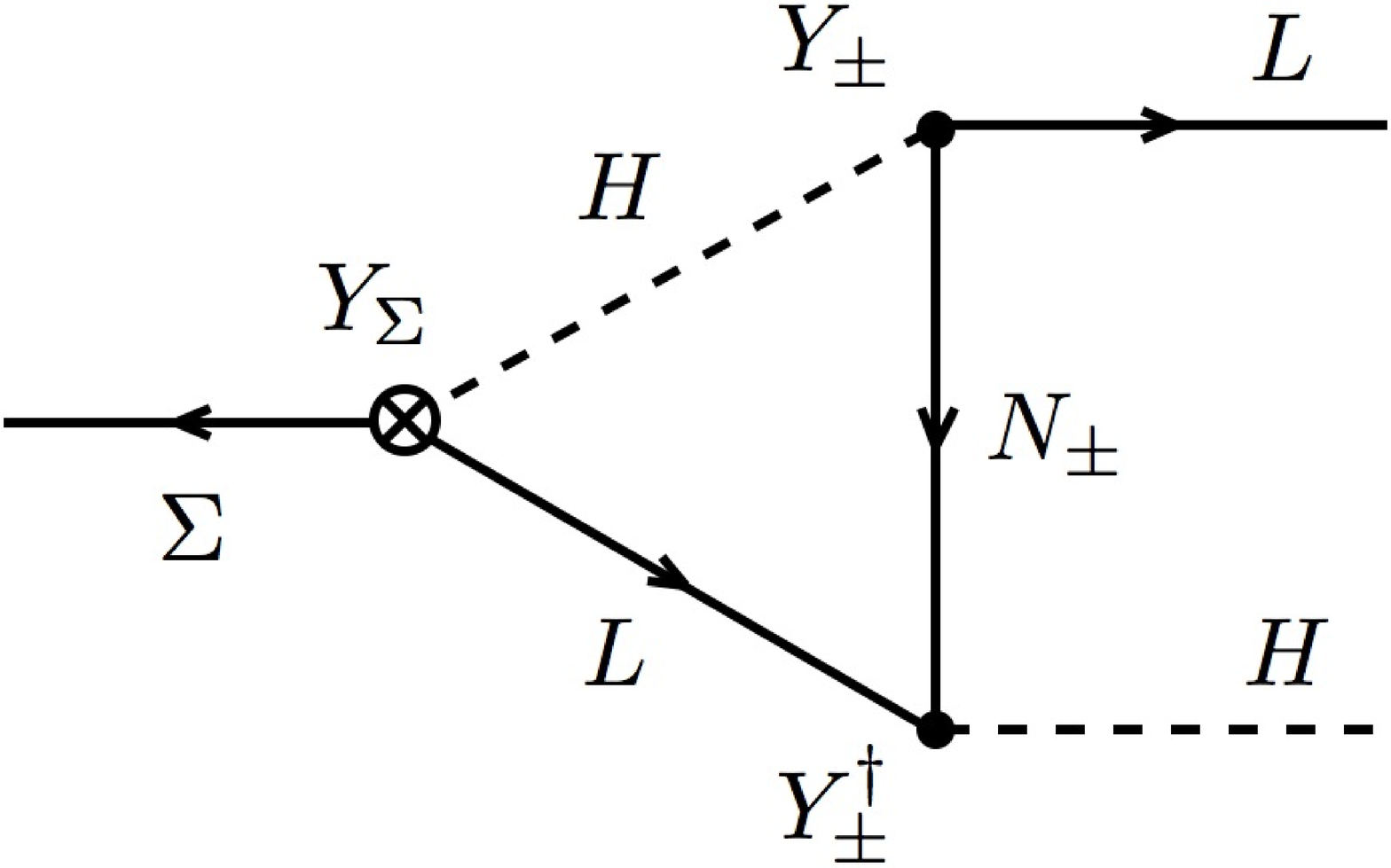}~~~~~~
\includegraphics[height=3cm]{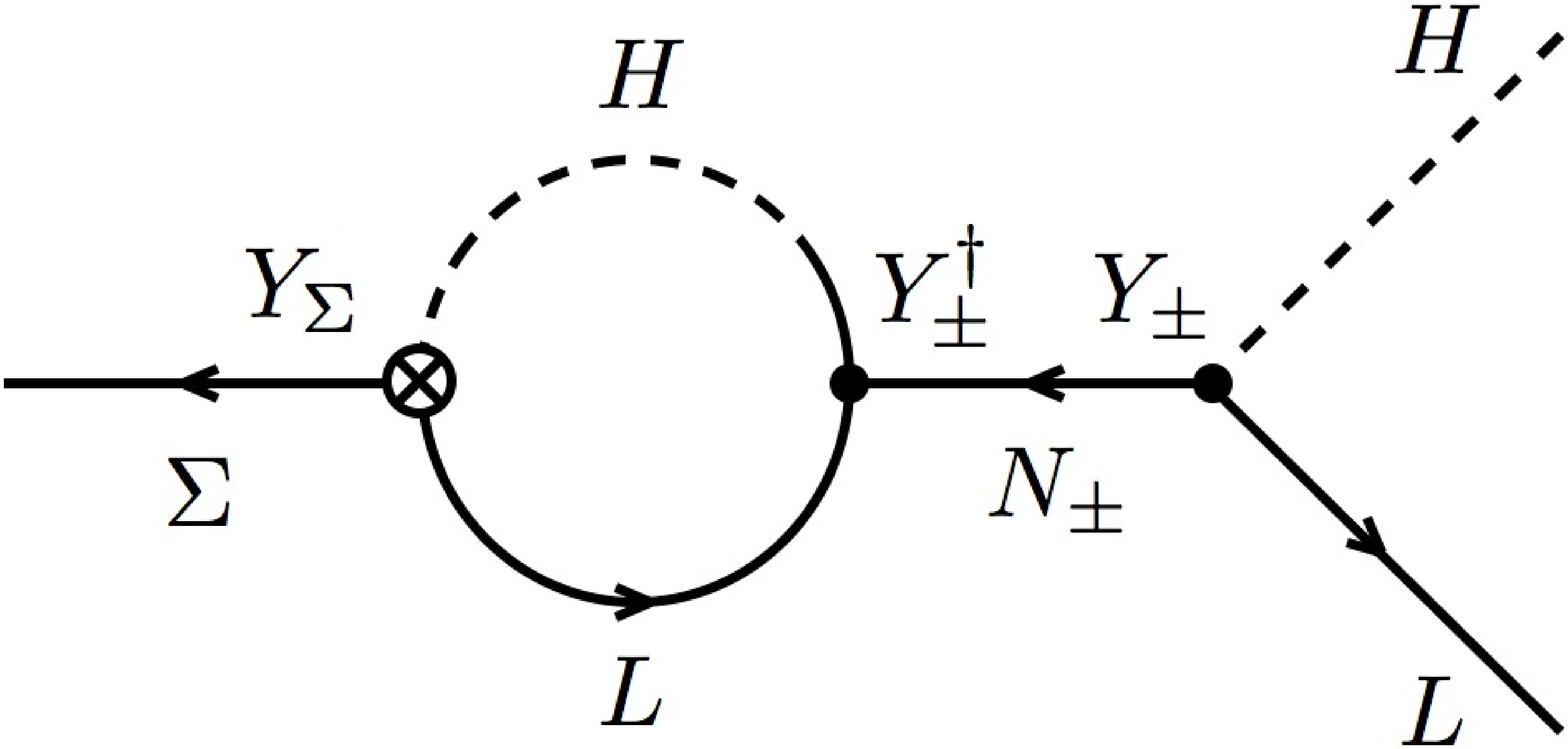}
\caption{\label{fig:l-g} Tree level and one loop diagrams for the
  decay of $\Sigma$ that interferes to generate a lepton asymmetry of
  the universe.}
\end{figure}

We now discuss the issue of leptogenesis in this model.  It can occur
only after the local $(B-L) \subset$ \10 symmetry is broken. This will
take place through the out-of-equilibrium decay of the singlet fermion
$\Sigma$.
The total width of $\Sigma$ is given by (treating $Y_{\Sigma}$ as a
column vector)
\begin{equation}
  \label{eq:wid}
\Gamma_{\Sigma}\sim \frac{1}{8\pi}Y_{\Sigma}^{\dagger}Y_{\Sigma} M_{\Sigma}  
\end{equation}
where $Y_{\Sigma}$ is given in Eq.~(\ref{eq:36}).

The asymmetry coming from the diagrams of Fig.~\ref{fig:l-g} involves
the sum over $k$ which reduces to the sum over the lightest pair of
the (almost degenerate) $N_{\pm}\equiv\frac{1}{\sqrt{2}}(\nu^{c}\pm
S)$ states with masses $\pm M_1$.  A comparison of the diagrams in
Fig.~\ref{fig:l-g} and Fig.~\ref{fig:tl+oneloop} gives the following
dictionary
\begin{equation}
  \label{eq:44}
  \begin{array}{cc}
   \lambda_{j 1} \rightarrow Y_{\Sigma j} &
  \lambda_{m 1} \rightarrow Y_{\Sigma m}\\
  \lambda_{m k} \rightarrow Y_{\pm m} &
  \lambda_{j k} \rightarrow Y_{\pm j}
  \end{array}
\end{equation}
We have then for the numerator 
\begin{eqnarray}
  \label{eq:45}
  \sum_{j,m,\pm} Y_{\Sigma j}^* Y_{\Sigma m}^* Y_{\pm m} Y_{\pm j}
  \frac{1}{M_{\pm}}&=&
 \frac{1}{M_1}\sum_{j,m} Y_{\Sigma j}^* Y_{\Sigma m}^* \left(Y_{+ m} Y_{+ j}
  -Y_{- m} Y_{- j} \right)\nonumber\\
&=& \frac{4}{M_1}\sum_{j,m} Y_{\Sigma j}^* Y_{\Sigma m}^* A_m B_j 
\end{eqnarray}
where we have defined:  $Y_{\pm m}= A_m \pm B_m$.  
%
%
Comparing with Eq.~(\ref{eq:36}) we get
\begin{equation}
  \label{eq:47}
  A_m = - \frac{1}{\sqrt{2}} 
\alpha_H \left(Y^T\, U_N\right)_{m1}, \qquad
 B_m= - \frac{1}{\sqrt{2}} \alpha_\chi^H 
\left( F\, U_S\right)_{m1} 
\end{equation}
Putting everything together we finally obtain
\begin{equation}
  \label{eq:asym}
\epsilon_{\Sigma}= -\frac{3}{8\pi}
\frac{M_{\Sigma}}{M_{1}}
\frac{{\rm Im}[(Y_{\Sigma}^{\dagger}
F_{k}U_{S}\alpha_{\chi^{k}}^H)_{1}(Y_{\Sigma}^{\dagger}
Y^T U_{N}\alpha_{H})_{1}]}{Y^{\dagger}_{\Sigma}Y_{\Sigma}}
\end{equation}
It is important to note that, in contrast to the asymmetry
$\epsilon_1$ in the minimal seesaw discussed in the previous section,
$\epsilon_{\Sigma}$ is not constrained by the light neutrino
masses. This can essentially be understood by the fact that the
neutrino mass, see eq.(\ref{eq:mnua}) and eq.(\ref{eq:vev-seesaw-2}),
is suppressed by the small ratio $v_L/v_R$, whereas in the calculation
of $\epsilon_{\Sigma}$, the small quantity $\Delta/v_R$ appears
quadratically in the numerator and the denominator, and thus
cancels. $\epsilon_{\Sigma}$ can therefore be much larger than in the
minimal seesaw case, independent of the light neutrino masses.
Consequently, there is also no {\em lower bound on $M_{\Sigma}$} from
the asymmetry parameter $\epsilon_{\Sigma}$.

\section{Ansatzes for the Coupling Matrices}
\label{sec:ansatz-coupl-matr}

We now estimate the resulting CP asymmetry needed for leptogenesis
making use of the current values of the neutrino oscillsation
parameters given in \cite{Maltoni:2004ei}.  A simple ansatz is to
assume that it comes just from the Dirac phase of the three-neutrino
lepton mixing matrix assuming the unitary approximation.  In this
approximation the asymmetry is proportional to the unique CP invariant
parameter that can be probed in neutrino oscillations.  The maximum
value of the asymmetry that can be achieved is obtained by varying
randomly all the other model parameters and the results are displayed
at each point of the plane ($\sin\theta_{13}$,$\sin\delta$).

\subsection{Generic case: non-hierarchical $Y_{ij}$}
\label{sec:nonhierarchicalY}

We start by considering a non-hierarchical ansatz for the coupling
matrices. In order to reduce the unknowns we assume that both
${F}_{ij}$ and $\tilde{F}_{ij}$ matrices are proportional to the
standard \10 Dirac Yukawa coupling matrix $Y_{ij}$, which is
symmetric, namely
\begin{equation}
  \label{eq:13}
  F_{ij}=f Y_{ij},\quad
  \tilde{F}_{ij}=\tilde{f} Y_{ij},\quad
  Y_{ij}=Y_{ji}
\end{equation}
With this choice Eq.~(\ref{eq:mnu}) can be written as,
\begin{equation}
  \label{eq:54}
 m_{\nu}=  m_{\nu}^{(a)}+ m_{\nu}^{(b)}
\end{equation}
where
\begin{eqnarray}
  \label{eq:12}
    m_{\nu}^{(a)}&=& - \frac{v^2}{\tilde{f}^2 v^2_R M_{\Sigma}}\, 
\Delta_i \Delta_j\simeq - 3\times 10^{10}\ \frac{1}{\tilde{f}^2}\
\frac{1}{\left(\frac{M_{\Sigma}}{10^3 \hbox{GeV}}\right)}\ 
\left( \frac{\Delta_i}{v_R}\right)
\left( \frac{\Delta_j}{v_R}\right)\  ({\rm eV})
\nonumber\\[+2mm]
  m_{\nu}^{(b)}&=& - \frac{f}{\tilde{f}}\ \frac{2 v^2}{M_X}\  Y_{ij}\simeq - 3\times 10^{-3}\,\frac{f}{\tilde{f}}\ Y_{ij}\  ({\rm eV})
\end{eqnarray}
Clearly, the projective nature of $m_{\nu}^{(a)}$ will not explain the
neutrino data, therefore we need the contribution of $m_{\nu}^{(b)}$.
The atmospheric scale can easily be reproduced with $Y\simeq
\mathcal{O}(1)$,~$f/\tilde{f} \simeq 0.1$ and $\frac{\Delta}{v_R}
\simeq 10^{-7}$, for $M_\Sigma = 1\ {\rm TeV}$.

With the assumptions of Eq.~(\ref{eq:13}) we can solve for the Yukawa
matrix in terms of the experimentally observed neutrino oscillation
parameters via
\begin{equation}
  \label{eq:14}
    Y_{ij}=-\frac{M_X}{2 v^2} \frac{\tilde{f}}{f}\left[  
\left(U^T\right)^{-1} m_{\nu}^{\rm exp}\ U^{-1} +
\Delta_i \Delta_j \frac{v^2}{v_R^2 M_{\Sigma}} \frac{1}{\tilde{f}^2}
\right]
\end{equation}
which also involves other parameters of the model.

With this ansatz and the choice $f=1,\ \tilde{f}=0.1$ we take the
lepton mixing parameters in the range allowed by the experiment.  At
$3\sigma$ the latest neutrino oscillation data
give~\cite{Maltoni:2004ei}
\begin{equation}
  \label{eq:48}
  \sin^2\theta_{12} \in [0.24,0.40],\quad  \sin^2\theta_{23}
 \in [0.34,0.68],\quad 
  \sin^3\theta_{13} < 0.04
\end{equation}
In addition we take the other model parameters in the following
ranges,
\begin{equation}
  \label{eq:11}
  M_{\Sigma} \in [10^3,10^7]\ {\rm GeV},\quad
  \Delta_i \in [10^{-3}, 10^3]\ {\rm GeV},\quad
  v_R \in [10^3,10^7]\ {\rm GeV}
\end{equation}

\begin{figure}[!h] \centering
 \includegraphics[clip,width=130mm]{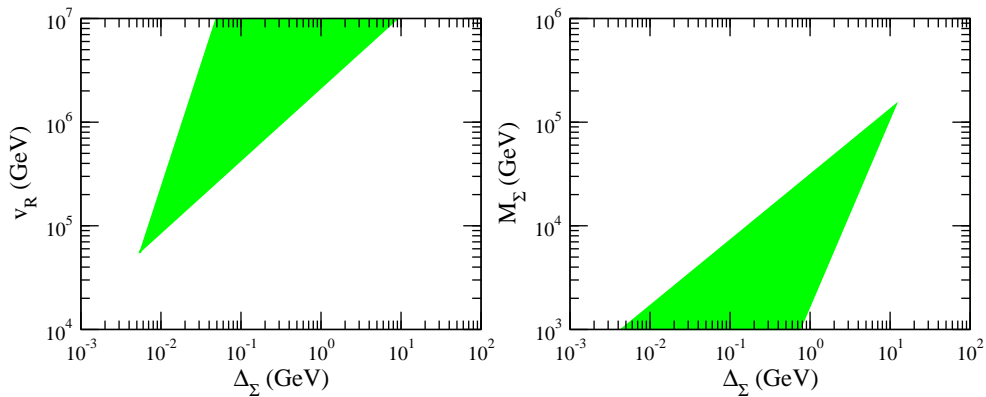}
 \vspace{-5mm}
 \caption{ $v_R$, $M_\Sigma$ and $\Delta_\Sigma$ values that follow
   from our first \textit{ansatz} Eq.~(\ref{eq:54}) for
   $\tilde{f}=0.1$, when the parameters are varied as in
   Eq.~(\ref{eq:11}), see text.}.
  \label{fig:Ureal}
\end{figure}

\begin{figure}
  \centering
  \includegraphics[height=0.25\textheight]{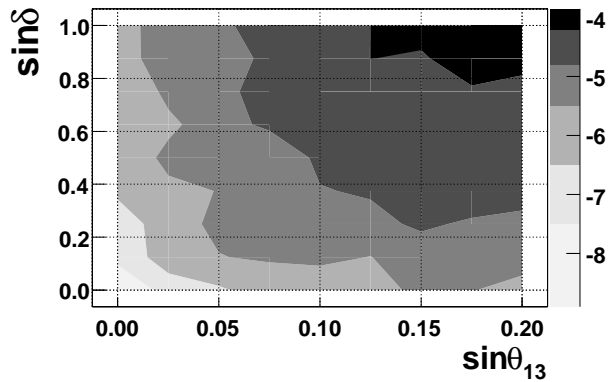} 
\caption{Contour levels of the logarithm of the maximum value of the asymmetry 
$\epsilon_\Sigma$ obtained as a function of $\sin\theta_{13}$ and $\sin\delta$
for the non--hierarchical ansatz of section IV A. The other neutrino
mixing parameters are taken the from the latest neutrino oscillation data
given in~\cite{Maltoni:2004ei}.}
 \label{fig:Ureal-2}
\end{figure}

\noindent
With these values we can see the resulting $v_R$, $M_\Sigma$ and
$\Delta_\Sigma$ values that follow from our ansatz in
Fig.~\ref{fig:Ureal}. They show that it is possible to fit the
neutrino data with this simple type of \textit{ansatz}.
The resulting CP asymmetry produced is given in
Fig.~\ref{fig:Ureal-2}.  Here we have calculated the maximum value of
the asymmetry that can be achieved at each point of the
($\sin\theta_{13}$,$\sin\delta$) plane, varying randomly all the other
parameters. Clearly the sizeable values of the asymmetry can be
obtained, especially for large $\sin\theta_{13}$ and $\sin\delta$, as
expected, so that the necessary CP asymmetry needed for leptogenesis
can be achieved.



One sees that large values of the asymmetry, compatible with the
experimental data, can be achieved in the full parameter space, even
for very small $\sin\theta_{13}$ and $\sin\delta$ values.
However the ansatz is manifestly inconsistent with the successful
Fritzsch texture for the quark masses.

\subsection{Fritzsch case: hierarchical $Y_{ij}$}
\label{sec:hirerarchicalY}

It is natural to ask whether our \10 model described by the Lagrangian
of Eq.~(\ref{eq:lag}) can provide thermal leptogenesis while
reconciling the successful Fritzsch texture for the quark masses with
the observed structure of lepton masses and mixings that follow from
neutrino oscillation experiments.
To this end we now assume that the $Y_{ij}$ Yukawa's involved in
neutrino mass generation are also restricted by the Fritzsch
\textit{ansatz} for the quark couplings, given in
Eq.~(\ref{eq:fritsch-1}), with $a,c$ complex and $b,d$ real. Aware of
the fact that the phases will be necessary in computing the final
asymmetry, we will, for the moment take the \textit{ansatz} parameters
all real, as
\begin{equation}
  \label{eq:fritsch-2}
  a=\sqrt{\frac{m_u m_c}{v^2}},\quad
  b=\frac{m_c}{v},\quad
  c=\sqrt{\frac{m_c m_t}{v^2}},\quad
  d=\frac{m_t}{v}
\end{equation}
These values imply a strong hierarchy among the Yukawa couplings. 
With this choice, let us now look at the neutrino mass matrix. It is
clear from Eq.~(\ref{eq:mnu}) that this hierarchy must be corrected by
a suitable hierarchy in the $\tilde{F}$ coupling matrix. In fact
Eq.~(\ref{eq:mnu}) can be ``solved'' for $\tilde{F}$ as
\begin{equation}
  \label{eq:53}
  \tilde{F}= X^{-1} F
\end{equation}
where $X$ is obtained from
\begin{equation}
  \label{eq:52}
  Y X^T + X Y^T = -\frac{M_X}{v^2}\left[ 
\left(U^T\right)^{-1} m_{\nu}^{\rm exp}\ U^{-1} +
\left(Y \tilde{\Delta}\right) \left(Y \tilde{\Delta}\right)^T
\frac{v^2}{v_R^2 M_{\Sigma}} \right] \equiv Z
\end{equation}
where $\tilde{\Delta}\equiv \tilde{F}^{-1} \Delta$,~$\tilde{F}$ being,
in general, an arbitrary non-symmetric matrix.
Now $Z$ is expressed as a combination of neutrino data and additional
parameters, for which we take random values. 

We have performed a random study of this Fritzsch ansatz assuming as
our ansatz that $\tilde{F} = X^{-1} F$ with $F=F^T$ and
taking random values of order one in various ways. We have found that
in this case, for example for $v_R$ in the range $[10^7,10^8]$ {\rm
  GeV}, $M_{\Sigma}$ of few TeV and $\Delta_\Sigma \in [10^{-2},10^4]$
{\rm GeV} one indeed obtains a viable solution.
One finds that some of the entries of $\tilde{F}$ are small (of order
$10^{-3}$) in order to compensate for the corresponding smallness of
$Y$ in Eq.~(\ref{eq:fritsch-1}).

\begin{figure}[!h] \centering
  \includegraphics[height=0.2\textheight]{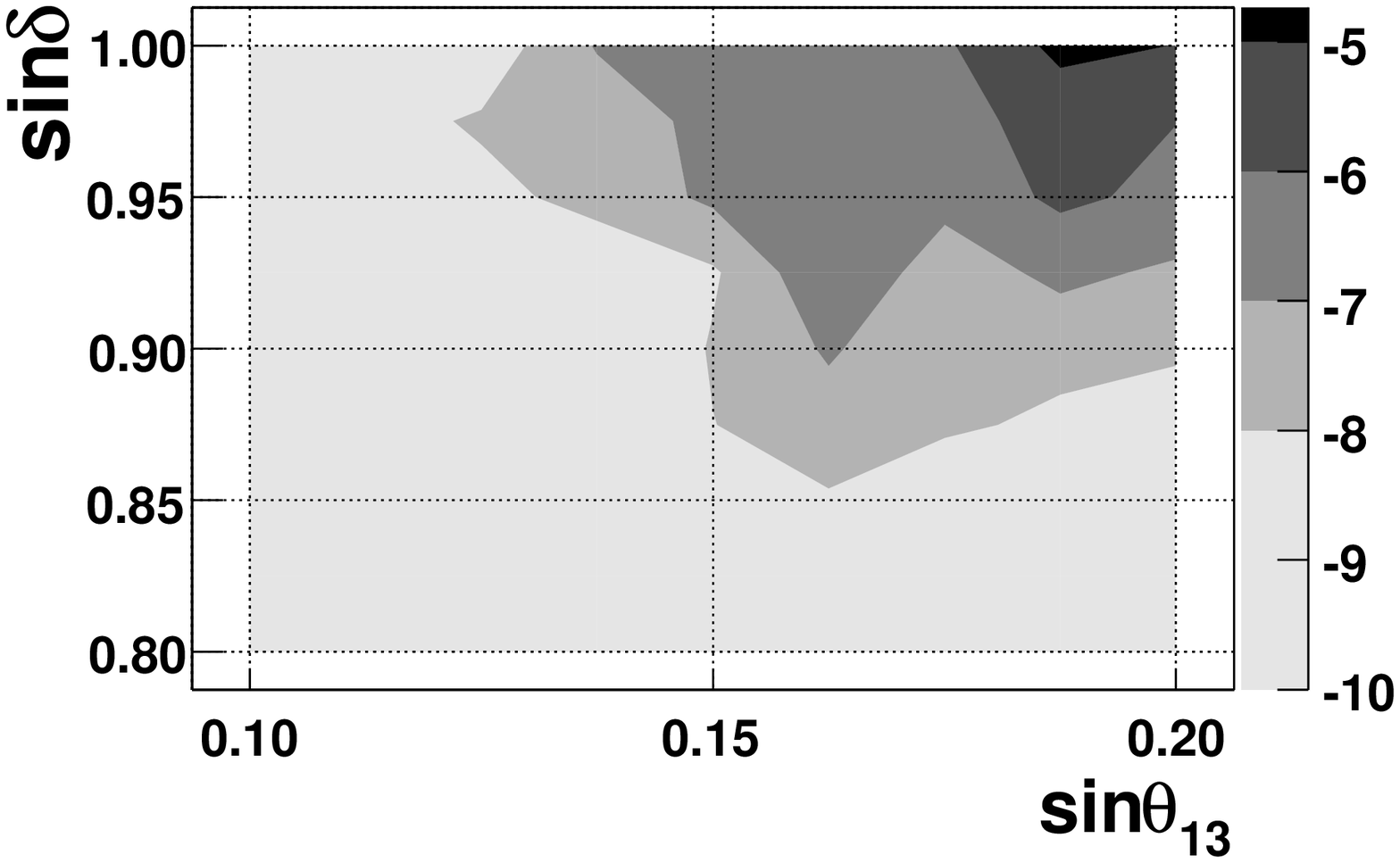}
  \includegraphics[height=0.2\textheight]{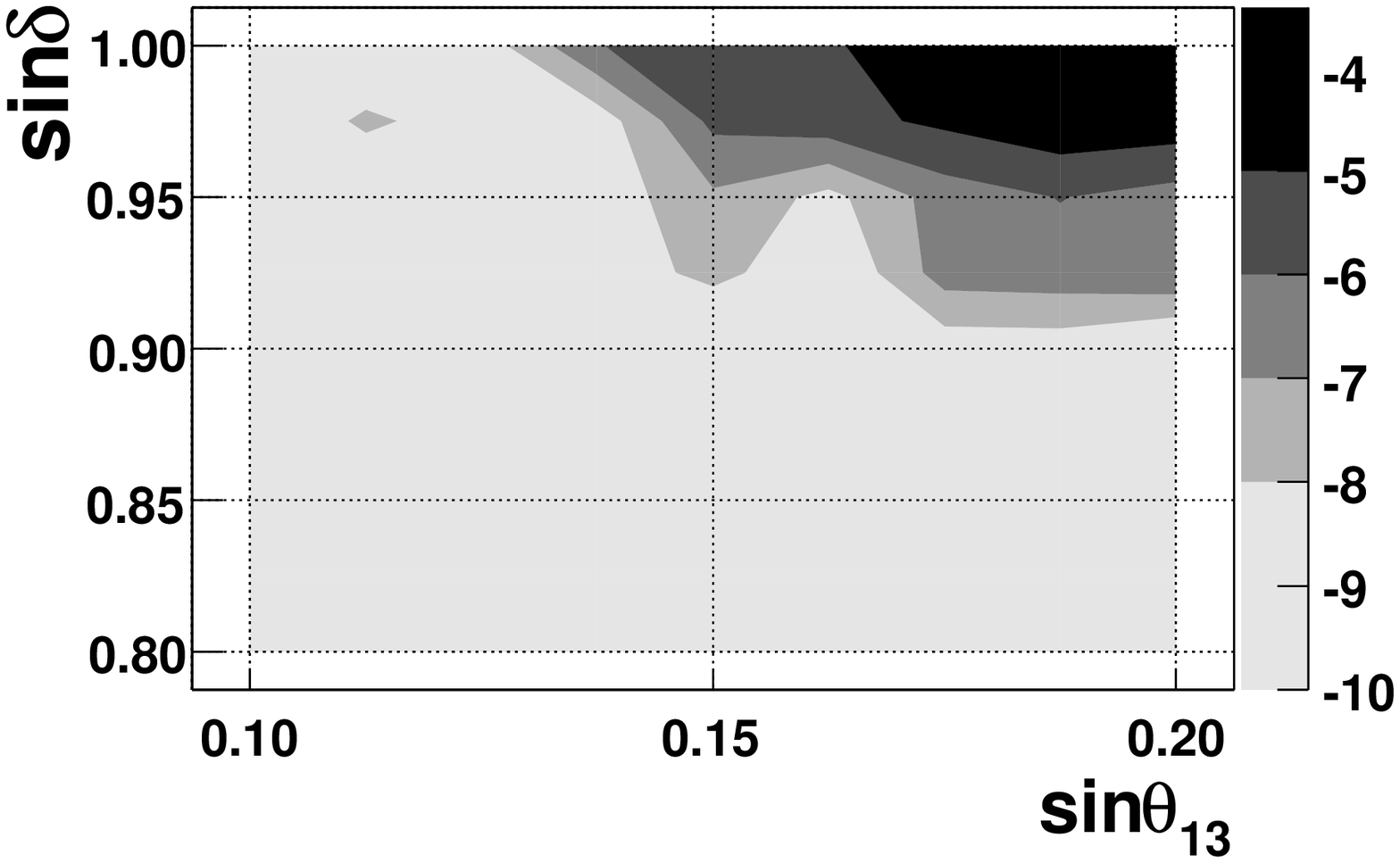}
  \caption{ Contour levels of the maximum asymmetry obtained as a
    function of $\sin\theta_{13}$ and $\sin\delta$ using the Fritzsch
    texture with $\tilde{F} = X^{-1} F$, F=I (left panel) or F
    symmetric (right panel).
\label{fig:small}}
\end{figure}

We have also explicitly calculated the value of the asymmetry
$\epsilon_{\Sigma}$ for this ansatz as explained above.
The results are shown in Fig.~\ref{fig:small} for the case of the
hierarchical ansatz with the current neutrino oscillation parameters
from \cite{Maltoni:2004ei} and the Dirac phase is $\delta \in
[0,\pi]$.

One sees that, in this case, values of the asymmetry of order
$10^{-6}$-$10^{-7}$ can only be obtained for large values of
$\sin\theta_{13}$ and $\sin\delta$.  Therefore, this scenario can be
potentially probed by the future neutrino oscillation measurements.

\begin{figure}[!h] \centering
  \includegraphics[height=0.2\textheight]{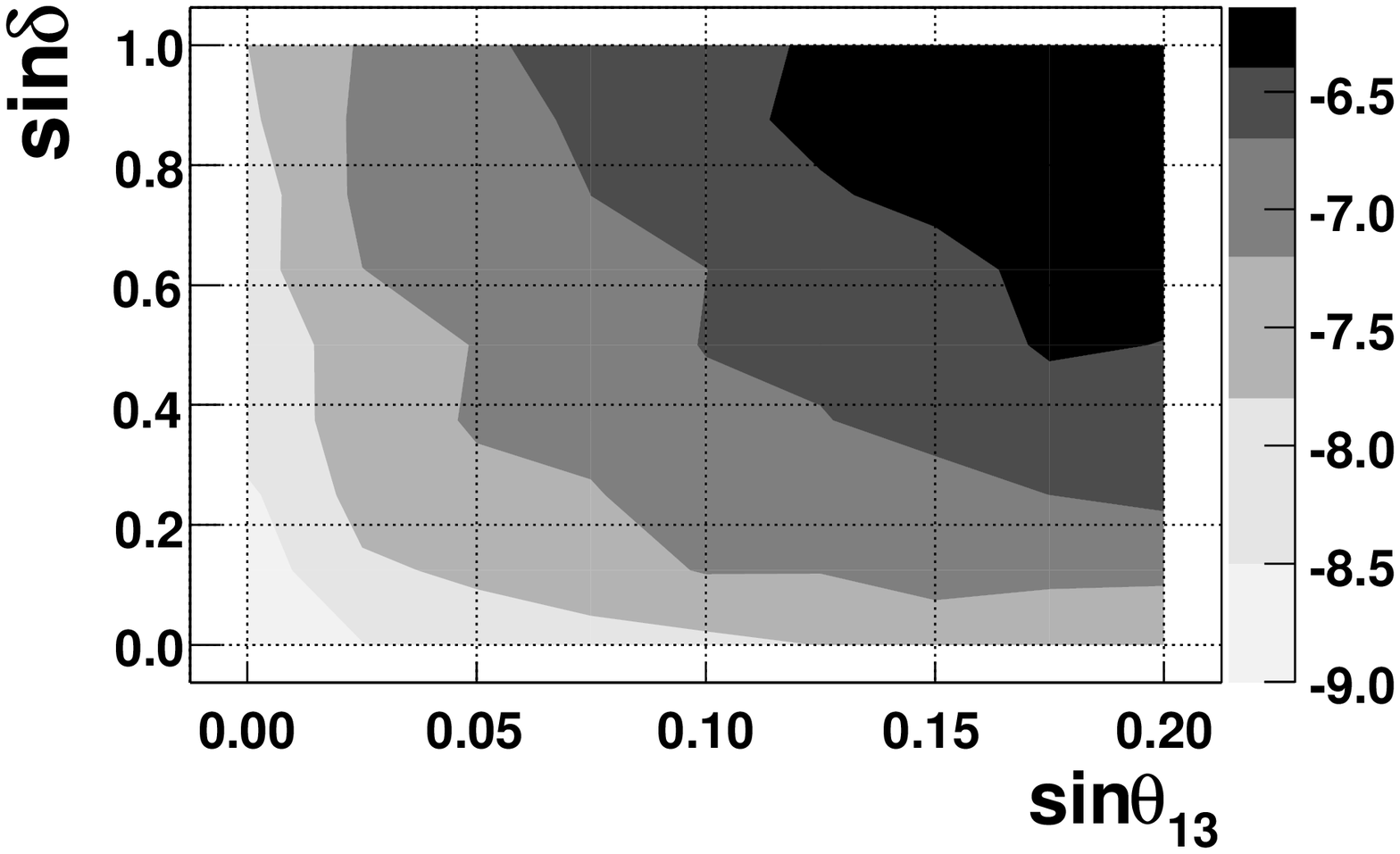}
  \includegraphics[height=0.2\textheight]{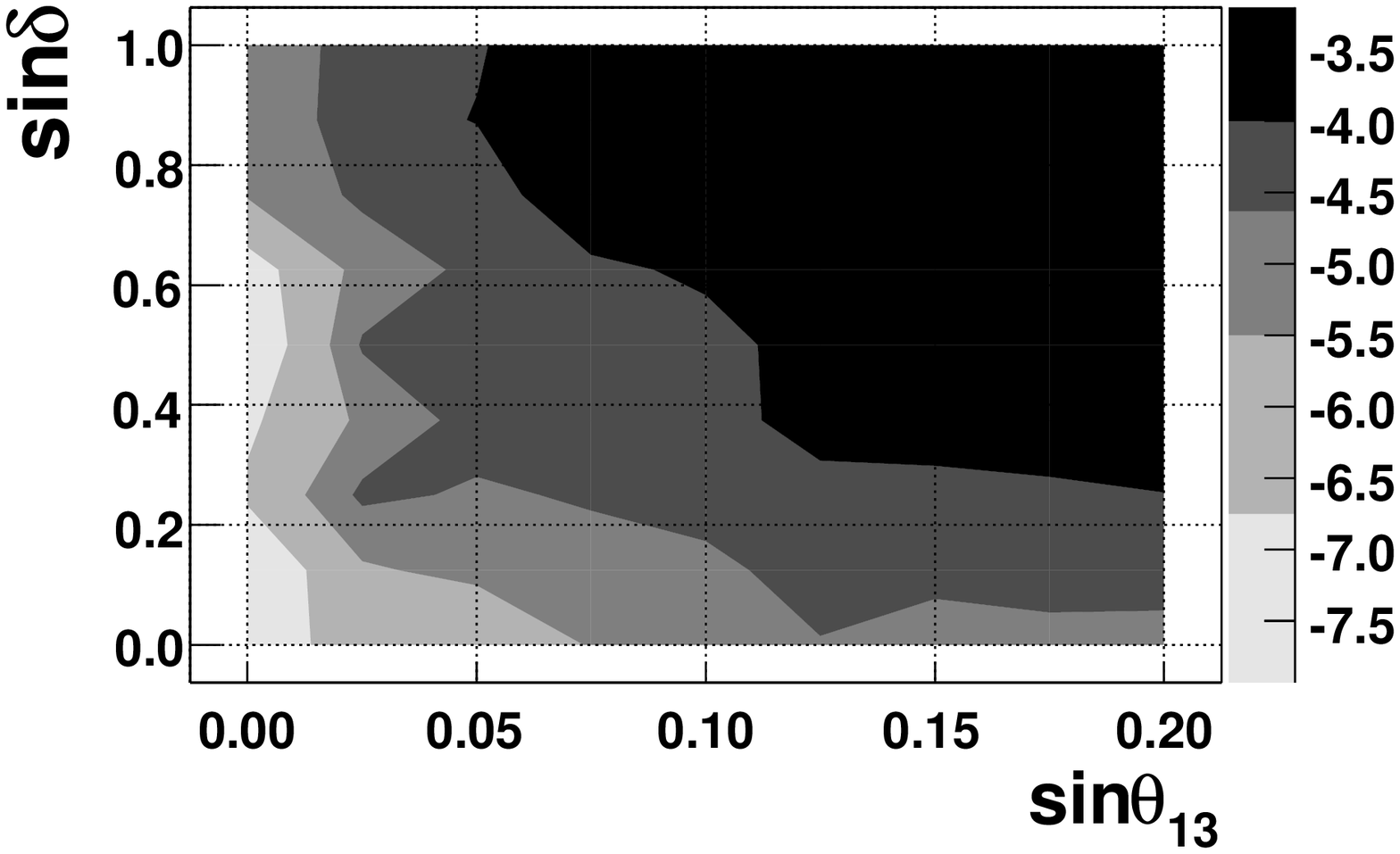}
  \caption{Contours of the maximum value of the asymmetry obtained
    using the Fritzsch texture. Left panel: $\tilde{F} = \tilde{f} Y$,
    $F$ symmetric. Right panel: $\tilde{F}$ diagonal and $F=X\tilde{F}$.
  }
\label{fig:larger}
\end{figure}

Larger values of the CP asymmetry compatible with current neutrino
oscillation measurements can be found for other ansatze of the current
model using the Fritzsch texture.  For instance, if we consider
$\tilde{F} = \tilde{f} Y$ with $F$ symmetric, or $\tilde{F}$ diagonal
and $F=X\tilde{F}$, one can obtain very large values of the asymmetry
even for very small values of $\sin\theta_{13}$ and $\sin\delta$, as
illustrated in Fig.~\ref{fig:larger}.

%

\section{Asymmetry washout, singlet production and the 
sphaleron constraint}
\label{sec:washout}

Generating a large enough asymmetry in the decay of the lightest
singlet is a necessary, but not a sufficient condition for successful
leptogenesis. 
Additional conditions are required, before one can conclude that any
given model can generate the required baryon asymmetry.

The first condition to be satisfied is the out-of-equilibrium decay of
the heavy singlet, which is nothing but the basic Sakharov condition
\cite{Sakharov:1967dj}. This requires that the decay rate is smaller
than the expansion rate of the universe, i.e.  $\Gamma_{\Sigma} \ll
\Gamma_{\rm Hubble}$ at the decay epoch. For large $\Gamma_{\Sigma}$,
say, of order of the expansion rate, part of the asymmetry produced in
the decays will be washed out by inverse scattering processes
violating lepton number. This constraint will thus put an upper limit
on the couplings of $\Sigma$ to the leptons (and Higgses) for any
given mass $M_{\Sigma}$. Second, we need to produce a sufficient
number of singlets in the early universe. Singlets could be either
produced by their couplings to the thermal bath or through the decay
of some heavier particle, which was present in the universe at earlier
times.  Obviously, a sufficiently strong direct thermal production
requires a minimum value for the couplings of $\Sigma$, whereas the
second option does not. 
The third constraint comes from the fact that the $\Sigma$ decays have
to take place at a time early enough that the SM sphalerons are still
in equilibrium, otherwise we will produce a lepton number, but no
non-zero baryon number. This constraint puts a lower limit on
$\Gamma_{\Sigma}$, $\Gamma_{\Sigma} \ge \Gamma_{\rm Sphaleron}$,
independent of the production mechanism of $\Sigma$.

To accurately calculate the first two of the above conditions, in
principle, one needs to set up a network of Boltzmann equations, which
in general can only be solved numerically \cite{Kolb:1990vq}. However,
under certain simplifying assumptions, one can derive approximate
analytical solutions which reproduce the full, numerical calculations
quite well. Several analytical approximations have been proposed in
the literature, see for example
\cite{Kolb:1990vq,Nielsen:2001fy,Buchmuller:2002rq}.  In our earlier
paper \cite{Hirsch:2006ft} we have used the following approximate form
for the washout factor:
\begin{equation}\label{defkapold}
\kappa'(z) = \frac{1}{1+10 z},
\end{equation}
where $z \equiv \frac{\Gamma}{\Gamma_{\rm Hubble}}$, with $\Gamma_{\rm
  Hubble}$ being the expansion rate of the universe
\cite{Kolb:1990vq}.  Ref. \cite{Buchmuller:2002rq} has numerically
solved the Boltzmann system for the case of the simpler type-I
seesaw. In this case the decay width of the lightest right-handed
neutrino is proportional to $\Gamma \sim (Y^{\dagger}Y)_{11}$, and the
authors of \cite{Buchmuller:2002rq} define the ``effective mass''
parameter ${\tilde m}_1 = (Y^{\dagger}Y)_{11} v^2/M_1$, where $M_1$ is
the mass of the lightest right-handed neutrino and $v$ the SM
vev. Thermal equilibrium for the right-handed neutrino is reached for
an ``equilibrium mass'' ${\tilde m}_{*} = 1.08 \cdot 10^{-3}$ eV, for
which by definition $z=1$. The full numerical calculation is then very
well approximated by
\begin{equation}\label{diBari}
\kappa(z) \simeq 0.24 (x_{-} e^{-x_{-}} + x_{+}e^{-x_{+}})
\end{equation}
with
\begin{equation}\label{defxmp}
x_{\pm} = \Big(\frac{{\tilde m}_{*}}{{\tilde m}_{\pm}}\Big)^{\mp 1 - \alpha},
\end{equation}
and the numerical values 
\begin{equation}
{\tilde m}_{+} = 8.3\times10^{-4} \hskip1mm {\rm eV}, \hskip5mm
{\tilde m}_{-} = 3.5\times10^{-4} \hskip1mm {\rm eV}, \hskip5mm
\alpha = 0.1.
\end{equation}
We can make use of the results of \cite{Buchmuller:2002rq} with the 
straightforward replacement,
\begin{equation}\label{newdefxmp}
x_{\pm} = \Big(\frac{{\tilde m}_{*}z}{{\tilde m}_{\pm}}\Big)^{\mp 1 - \alpha}.
\end{equation}
Our more complicated setup of Boltzmann equations then is solved
approximately by the fit eq. (\ref{diBari}). Note, that this
implicitly assumes \footnote{The same assumption is used in the fit
leading to this equation in the simpler type-I seesaw} that the
heavier singlets are sufficiently decoupled so as to not contribute
significantly to the washout.

\begin{figure}[!h] \centering
  \includegraphics[height=70mm,width=100mm]{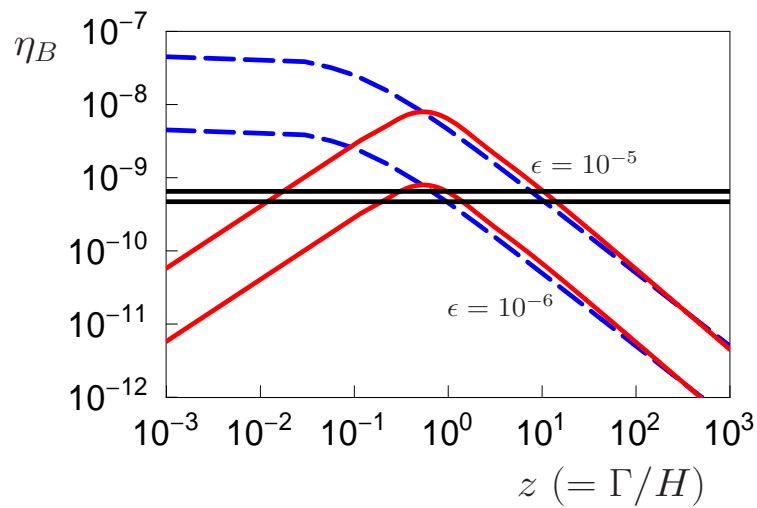}
  \caption{Calculated range for $\eta_B$ as a function of $z$, for the 
different fits to the washout factor. The dashed (full) lines show the 
approximate analytical solutions to the Boltzmann equations, see 
eqs (\ref{defkapold}) and (\ref{diBari}). Results are shown 
for 2 typical values of $\epsilon=10^{-5},10^{-6}$. The horizontal 
black band is the currently allowed  $\eta_B$ range \cite{pdg}.}
\label{fig:etaBtypeI}
\end{figure}

As seen in Fig. (\ref{fig:etaBtypeI}) eq. (\ref{diBari}) and
eq. (\ref{defkapold}) lead to very similar results for $z > 1$,
The two forms differ, however, for $z<1$. This can be traced to the
fact that eq. (\ref{diBari}) also accounts for the suppressed
production in the weak coupling regime, while eq. (\ref{defkapold})
does not. For definiteness, in the plots shown below, we have used
eq. (\ref{diBari}).

We have also used the constraint imposed by the condition that the 
decay of the singlets must happen while the sphalerons are still 
active. We estimate the sphaleron time using
\begin{equation}\label{est_Sphal}
\Gamma_{\rm Sphaleron} = \Gamma_{\rm Hubble}|_{T=EWPT} 
 = \sqrt{4\pi^3 g_{*}/45}(\frac{T^2}{M_{Pl}})|_{T=EWPT},
\end{equation}
with $EWPT$ denoting the energy at which the electro-weak phase
transition occurs, $g_{*}$ the effective number of degrees of freedom
and $M_{Pl}$ the reduced Planck mass. We cut all points with
$\Gamma_{\Sigma} \le \Gamma_{\rm Sphaleron}$. This is, of course, a
rough approximation to the real, dynamical situation. However, we
believe it to be sufficiently accurate for deriving order of magnitude
constraints on the model parameters. Finally, when converting the
produced lepton asymmetry to the baryon asymmetry, we have to take
into account an efficiency factor for the sphalerons.  This factor has
been calculated in \cite{Khlebnikov:1988sr} to be
\begin{equation}\label{sp_eff}
\eta_B = \frac{8 n_F + 4 n_H}{22 n_F + 13 n_H} \eta_L,
\end{equation}
where $n_F$ ($n_H$) is the number of families (Higgses). Numerically 
this factor is $\sim 1/3$. 

\begin{figure}[!h] \centering
  \includegraphics[height=70mm]{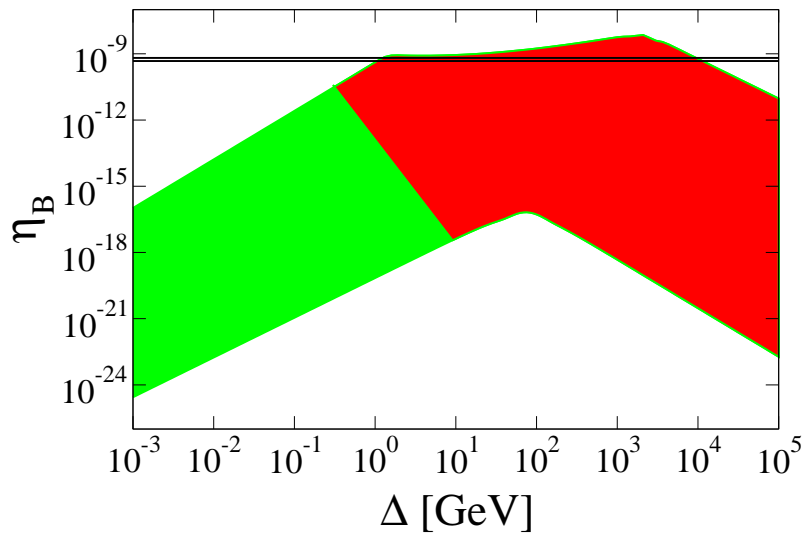}
  \caption{Calculated range for $\eta_B$ as a function of $\Delta$,
for the ansatz defined in fig. (\ref{fig:small}), to the left.  The
horizontal band is the allowed range of $\eta_B$.  The full shaded
region corresponds to the allowed range without the sphaleron
constraint, while the superimposed darker/red area indicates the
allowed range of parameters once the sphaleron constraint is
included. For a discussion see text.}
\label{fig:etaB}
\end{figure}

Fig. (\ref{fig:etaB}) shows as an example the resulting $\eta_B$ as a
function of $\Delta$ for a numerical scan using the ansatz shown in
fig.(\ref{fig:small}), to the left. The light/dark (green/red) area is
the calculated range for $\eta_B$ without/with the sphaleron
constraint. One sees that the different constraints discussed above
conspire to choose a rather well-defined allowed range for the
parameter $\Delta$ in this case. A large enough baryon asymmetry can
be obtained roughly for $\Delta = [1,10^4]$ GeV, for the range 
of the other parameters as given in eq. (\ref{eq:11}).  It is amusing to
note, that the requirement of producing a sufficient number of
singlets and the sphaleron constraint lead to rather similar lower
cuts on $\Delta$ \footnote{The same will occur for the seesaw type-I:
The sphaleron constraints provides a lower cut on the ``effective mass
parameter'' ${\tilde m}_{1}$.}.

\begin{figure}[!h] \centering
   \includegraphics[height=60mm]{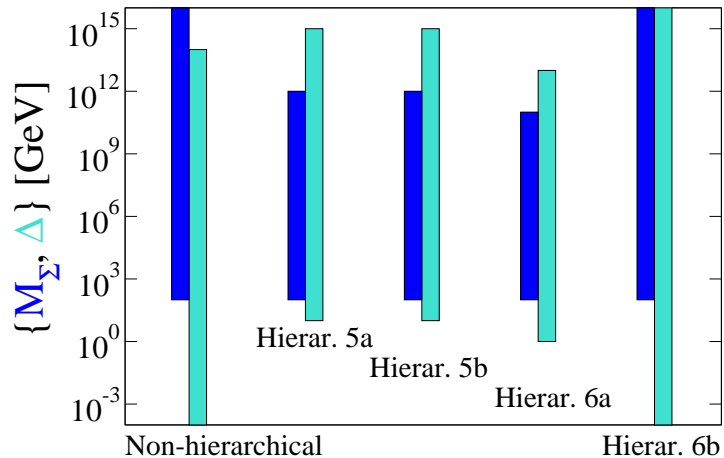}
   \caption{Calculated ranges for $\Delta$ and $M_{\Sigma}$ for the 
 different texture ans\"atze discussed in the text. This ranges have 
been obtained with the largest possible scan for the parameters, for 
example, $M_{\Sigma}$ in the full interval [$10^2,10^{16}$] GeV.} 
 \label{fig:alldelta}
 \end{figure}

We have repeated this exercise for all the different ans\"atze
discussed above. The resulting allowed ranges for the parameters
$\Delta$ and $M_{\Sigma}$ are shown in fig. (\ref{fig:alldelta}).
As shown in this figure, the random non-hierarchical ansatz and the
hierarchical ansatz shown in the right panel of Fig.~\ref{fig:larger}
lead to the widest allowed ranges for $\Delta$ and $M_{\Sigma}$. 

\begin{table}[t] \centering
  
  \begin{tabular}{|cccccc|}
    \hline
    $\Delta$[GeV]  & $M_\Sigma$[GeV] & $M_1$[GeV]  & $M_2$[GeV] & $M_3$[GeV] & $\eta_B$ [$10^{-10}$] 
    \\
    \hline
    \multicolumn{6}{|c|}{Non-hierarchical} \\
    \hline
    8.3$\times10^{-2}$  & 1.2$\times10^3$  & 4.1$\times10^4$  & 5.0$\times10^4$  & 2.6$\times10^5$  & 6.3 \\
    9.8  & 2.3$\times10^4$  & 9.3$\times10^5$  & 1.2$\times10^6$  & 6.1$\times10^6$  & 5.0 \\
    \hline
    \multicolumn{6}{|c|}{Hierarchical 5a} \\
    \hline
    3.5$\times10^3$ & 1.2$\times10^3$ & 3.6$\times10^4$ & 6.2$\times10^4$ &  6.7$\times10^6$  & 4.9 \\
    2.9$\times10^3$ & 1.1$\times10^3$ & 4.6$\times10^4$ & 6.8$\times10^4$ &  5.7$\times10^6$  & 5.8 \\
    \hline
    \multicolumn{6}{|c|}{Hierarchical 5b} \\
    \hline
    1.8$\times10^3$ & 6.6$\times10^2$ & 2.1$\times10^4$ & 6.4$\times10^4$ &  3.8$\times10^6$  & 5.2 \\
    8.3$\times10^2$ & 2.5$\times10^2$ & 6.1$\times10^3$ & 6.2$\times10^4$ &  9.8$\times10^6$  & 4.8 \\
     \hline
    \multicolumn{6}{|c|}{Hierarchical 6a} \\
    \hline
   2.9$\times10^2$ & 2.9$\times10^3$ & 1.6$\times10^5$ & 3.6$\times10^9$ &  6.0$\times10^{11}$ & 5.1 \\
   3.3$\times10^3$ & 8.3$\times10^2$ & 2.5$\times10^5$ & 1.3$\times10^{11}$ &  1.4$\times10^{13}$ & 5.8  \\
    \hline
    \multicolumn{6}{|c|}{Hierarchical 6b} \\
    \hline
    1.0$\times10^{1}$  & 1.8$\times10^3$ & 1.9$\times10^5$ & 1.1$\times10^6$ &  2.0$\times10^6$ &  5.4 \\
    3.4$\times10^{1}$  & 2.8$\times10^3$ & 1.5$\times10^6$ & 1.5$\times10^6$ &  2.5$\times10^6$ &  5.9 \\
    \hline
  \end{tabular}
  \caption{ \label{tab:example} 
Some example spectra for the different ans\"atze considered.}
\end{table}

Before we close this section let us briefly illustrate possible values
for the relevant parameters. These are $\Delta,~M_\Sigma$, the masses
$M_1,~M_2,~M_3$ of the three iso-singlet neutrinos and the resulting
baryon asymmetry $\eta_B$. Some examples are given in table~\ref{tab:example}
for each of the scenarios we have discussed. As one can see the 
acceptable baryon asymmetry may arise for rather low scales, in contrast 
to the (simpler) seesaw type-I schemes. Note, however that the values 
given in the table are by no means unique, and are also not meant to 
be ``representative'' of the classes. The generation of the baryon 
asymmetry is far easier in this context than in the traditionl seesaw 
model. The presence of the new singlets allows us to have, in addition, 
acceptable textures for the quark and lepton mixing angles.

\section{Conclusions}

We have reviewed the argument that in minimal type-I \10 seesaw one
can not easily reconcile the thermal leptogenesis scenario with the
successful Fritzsch texture for the quarks and an acceptable pattern
of lepton masses and mixings that follow from neutrino oscillation
experiments.
This is due to the fact that the large seesaw scale needed to account
for small neutrino masses leads to an overproduction of cosmological
gravitinos, which destroys the standard predictions of Big Bang
Nucleosynthesis (BBN). Barring the very special case of resonant
leptogenesis, one must go beyond the minimal type-I seesaw mechanism.

In this paper we have studied in some detail an extended seesaw scenario 
as a natural way to overcome this limitation. The proposed extension has 
the added virtue of providing a natural setting for reconciling the observed 
structure of neutrino mixing angles with the strong hierarchy among quark 
masses and the smallness of the quark mixings.
While this can always be accommodated in a ``generic'' unified theory
with arbitrary multiplet content, it becomes a real challenge for
unified predictive models of flavour.

We have provided a quantitative study of fermion masses and (an 
approximative calculation of) leptogenesis in the context of supersymmetric 
\10 Unification.  Our approach was phenomenological in that we have not 
assumed a specific flavour symmetry. We have shown how thermal leptogenesis 
can occur at relatively low scale through the decay of a new singlet, thereby
avoiding the gravitino crisis.  Washout of the asymmetry is
effectively suppressed by the absence of direct couplings of the
singlet to leptons.  For illustration we have shown how one can
accommodate current oscillation neutrino data and the required value
for the asymmetry for successful leptogenesis even if the only source
of CP violation is the Dirac phase $\delta$ in the low energy neutrino
mixing matrix. Finally, we note that we have not taken into account flavour 
effects \cite{Abada:2006ea,Nardi:2006fx} in the calculation of the asymmetry. 
However, we believe that the absence of a lower bound on $M_{\Sigma}$ 
in our model is independent on whether flavour effects are taken 
into account or not.

Using the Fritzsch texture we have found that some ansatze lead to
acceptable values of the asymmetry of order $10^{-6}$-$10^{-7}$ only
for large values of $\sin\theta_{13}$ and $\sin\delta$.  Therefore,
these scenarios can be potentially probed by the future neutrino
oscillation measurements, as the required value of the CP invariant is
nearly maximal.
In contrast, we have also presented alternative Fritzsch-type ansatze
leading to substantially larger values of the CP asymmetry, even for
very small values of $\sin\theta_{13}$ and $\sin\delta$. 
We have also discussed, how to approximately treat the conversion of
the decay asymmetry to the baryon asymmetry, without resorting to a
full numerical solution of the Boltzmann equations. To this end, we
made use of some approximation formulas derived previously for seesaw
type-I and discussed how they can be adapted to cover also our more
complicated case.

In summary, our extended seesaw scenario provides a way of reconciling
the lepton and quark mixing angles with thernal leptogenesis in a
unified scenario.
While this by itself does not constitute a complete theory of fermion
masses and leptogenesis, at least it provides a useful first step
towards an ultimate unified theory incorporating flavour.

\section*{Acknowledgments}

Work supported by MEC grants FPA2005-01269 and FPA2005-25348-E, by
Generalitat Valenciana ACOMP06/154, by European Commission Contracts
MRTN-CT-2004-503369 and ILIAS/N6 WP1 RII3-CT-2004-506222.


\begin{thebibliography}{10}

\bibitem{Maltoni:2004ei} For an updated review see M.~Maltoni,
  T.~Schwetz, M.~A. Tortola, and J.~W.~F. Valle, \newblock New J.
  Phys. {\bf 6}, 122 (2004), \newblock hep-ph/0405172 (v6) provides
  updated results as of September 2007; previous works by other groups
  are referenced therein and in~\cite{Fogli:2005cq}.

\bibitem{Fogli:2005cq}
 G.~L.~Fogli, E.~Lisi, A.~Marrone and A.~Palazzo,
  Prog.\ Part.\ Nucl.\ Phys.\  {\bf 57} (2006) 742
  [hep-ph/0506083].

\bibitem{Babu:1998wi}
K.~S. Babu, J.~C. Pati, and F.~Wilczek,
\newblock Nucl. Phys. {\bf B566}, 33 (2000), hep-ph/9812538.

\bibitem{bertolini2004fms}
S.~Bertolini, M.~Frigerio, and M.~Malinsk\'{y},
\newblock Physical Review D {\bf 70}, 095002 (2004).

\bibitem{Dermisek:2005ij}
R.~Dermisek and S.~Raby,
\newblock Phys. Lett. {\bf B622}, 327 (2005), hep-ph/0507045.

\bibitem{Altarelli:2004za}
G.~Altarelli and F.~Feruglio,
\newblock New J. Phys. {\bf 6}, 106 (2004), hep-ph/0405048.

\bibitem{Nasri:2004rm}
S.~Nasri, J.~Schechter, and S.~Moussa,
\newblock Phys. Rev. {\bf D70}, 053005 (2004), hep-ph/0402176.

\bibitem{schechter:1980gr}
J.~Schechter and J.~W.~F. Valle,
\newblock Phys. Rev. {\bf D22}, 2227 (1980).

\bibitem{Fukugita:1986hr}
M.~Fukugita and T.~Yanagida,
\newblock Phys. Lett. {\bf B174}, 45 (1986).

\bibitem{Khlopov:1984pf}
M.~Y. Khlopov and A.~D. Linde,
\newblock Phys. Lett. {\bf B138}, 265 (1984).

\bibitem{Ellis:1984eq}
J.~R. Ellis, J.~E. Kim, and D.~V. Nanopoulos,
\newblock Phys. Lett. {\bf B145}, 181 (1984).

\bibitem{Kawasaki:2004qu}
M.~Kawasaki, K.~Kohri, and T.~Moroi,
\newblock Phys. Rev. {\bf D71}, 083502 (2005), astro-ph/0408426.

\bibitem{Pilaftsis:2005rv}
A.~Pilaftsis and T.~E.~J. Underwood,
\newblock Phys. Rev. {\bf D72}, 113001 (2005), hep-ph/0506107.

\bibitem{Akhmedov:2003dg}
E.~K. Akhmedov, M.~Frigerio, and A.~Y. Smirnov,
\newblock JHEP {\bf 09}, 021 (2003), hep-ph/0305322.

\bibitem{Farzan:2005ez}
Y.~Farzan and J.~W.~F. Valle,
\newblock Phys. Rev. Lett. {\bf 96}, 011601 (2006), hep-ph/0509280.

\bibitem{Hirsch:2006ft}
M.~Hirsch, J.~W.~F. Valle, M.~Malinsky, J.~C. Romao, and U.~Sarkar,
\newblock Phys. Rev. {\bf D75}, 011701 (2007), hep-ph/0608006.

\bibitem{Malinsky:2005bi}
M.~Malinsky, J.~C. Romao, and J.~W.~F. Valle,
\newblock Phys. Rev. Lett. {\bf 95}, 161801 (2005), hep-ph/0506296.

\bibitem{Fritzsch:1977vd}
H.~Fritzsch,
\newblock Phys. Lett. {\bf B73}, 317 (1978).

\bibitem{babu:2002dz}
K.~S. Babu, E.~Ma, and J.~W.~F. Valle,
\newblock Phys. Lett. {\bf B552}, 207 (2003), hep-ph/0206292.

\bibitem{Altarelli:2005yx}
G.~Altarelli and F.~Feruglio,
\newblock Nucl. Phys. {\bf B741}, 215 (2006), hep-ph/0512103.

\bibitem{Hirsch:2006je}
M.~Hirsch, E.~Ma, J.~C. Romao, J.~W.~F. Valle, and A.~Villanova~del Moral,
\newblock Phys.\ Rev.\  D {\bf 75}, 053006 (2007), hep-ph/0606082.

\bibitem{Hirsch:2005mc}
M.~Hirsch, A.~Villanova~del Moral, J.~W.~F. Valle, and E.~Ma,
\newblock Phys. Rev. {\bf D72}, 091301 (2005), hep-ph/0507148.

\bibitem{Hirsch:2007kh}
M.~Hirsch, A.~S. Joshipura, S.~Kaneko, and J.~W.~F. Valle,
\newblock Phys.\ Rev.\ Lett.\  {\bf 99}, 151802 (2007), hep-ph/0703046.

\bibitem{Valle:2006vb}
  For an updated seesaw review see J.~W.~F. Valle,
\newblock J. Phys. Conf. Ser. {\bf 53}, 473 (2006), hep-ph/0608101,
\newblock Review based on lectures at the Corfu Summer Institute on Elementary
  Particle Physics in September 2005.

\bibitem{Minkowski:1977sc}
P.~Minkowski,
\newblock Phys. Lett. {\bf B67}, 421 (1977).

\bibitem{schechter:1982cv}
J.~Schechter and J.~W.~F. Valle,
\newblock Phys. Rev. {\bf D25}, 774 (1982).

\bibitem{covi:1996wh}
L.~Covi, E.~Roulet, and F.~Vissani,
\newblock Phys. Lett. {\bf B384}, 169 (1996), hep-ph/9605319.

\bibitem{Buchmuller:2004nz}
W.~Buchmuller, P.~Di~Bari, and M.~Plumacher,
\newblock Ann. Phys. {\bf 315}, 305 (2005), hep-ph/0401240.

\bibitem{Davidson:2002qv}
  S.~Davidson and A.~Ibarra,
  Phys.\ Lett.\  B {\bf 535} (2002) 25
  [arXiv:hep-ph/0202239].

\bibitem{Hamaguchi:2001gw}
  K.~Hamaguchi, H.~Murayama and T.~Yanagida,
  Phys.\ Rev.\  D {\bf 65} (2002) 043512
  [arXiv:hep-ph/0109030].

\bibitem{Barbieri:1999ma}
  R.~Barbieri, P.~Creminelli, A.~Strumia and N.~Tetradis,
  Nucl.\ Phys.\  B {\bf 575} (2000) 61
  [arXiv:hep-ph/9911315].

\bibitem{Sakharov:1967dj}
A.~D.~Sakharov,
Pisma Zh.\ Eksp.\ Teor.\ Fiz.\  {\bf 5} (1967) 32
[JETP Lett.\  {\bf 5} (1967\ SOPUA,34,392-393.1991\ UFNAA,161,61-64.1991) 24].

\bibitem{Kolb:1990vq}
  E.~W.~Kolb and M.~S.~Turner,
  Front.\ Phys.\  {\bf 69} (1990) 1.

\bibitem{Nielsen:2001fy}
  H.~B.~Nielsen and Y.~Takanishi,
  Phys.\ Lett.\  B {\bf 507} (2001) 241
  [arXiv:hep-ph/0101307].

\bibitem{Buchmuller:2002rq}
  W.~Buchmuller, P.~Di Bari and M.~Plumacher,
  Nucl.\ Phys.\  B {\bf 643} (2002) 367
  [arXiv:hep-ph/0205349].

\bibitem{Khlebnikov:1988sr}
  S.~Y.~Khlebnikov and M.~E.~Shaposhnikov,
  Nucl.\ Phys.\  B {\bf 308} (1988) 885.

\bibitem{pdg}
W.-M. Yao et al., J. Phys. G 33, 1 (2006), Particle Data Group, 
http://pdg.lbl.gov/ 

\bibitem{Abada:2006ea}
A.~Abada, S.~Davidson, A.~Ibarra, F.~X.~Josse-Michaux, M.~Losada and A.~Riotto,
JHEP {\bf 0609}, 010 (2006)
[arXiv:hep-ph/0605281].

\bibitem{Nardi:2006fx}
  E.~Nardi, Y.~Nir, E.~Roulet and J.~Racker,
  JHEP {\bf 0601}, 164 (2006)
  [arXiv:hep-ph/0601084].

\end{thebibliography}

\end{document}